\documentclass[preprint]{elsarticle}
\usepackage{natbib}
\usepackage{amsmath,amssymb,dsfont}
\usepackage{url}
\usepackage{xcolor}
\usepackage{enumerate}
\usepackage{graphicx}
\usepackage{placeins}
\renewcommand{\leq}{\leqslant}
\renewcommand{\geq}{\geqslant}
\renewcommand{\P}{\mathbb{P}}
\newtheorem{theo}{Theorem}
\newdefinition{df}{Definition}
\newdefinition{rmk}{Remark}
\newcommand{\N}{\mathbb{N}}
\newcommand{\R}{\mathbb{R}}
\newcommand{\eref}[1]{(\ref{#1})}
\newcommand{\E}{\mathbb{E}}
\newcommand{\V}{\mathbb{V}}
\newcommand{\cB}{\mathcal{B}}
\newcommand{\cN}{\mathcal{N}}
\newcommand{\cP}{\mathcal{P}}
\newcommand{\Z}{\mathcal{Z}}
\newcommand{\U}{\mathcal{U}}
\newcommand{\cS}{\mathcal{S}}
\newcommand{\T}{\mathcal{T}}

\begin{document}
\title{Testing $k$-monotonicity of a discrete
  distribution. Application to the estimation of the number of classes
  in a population}
\author[orsay]{J.~Giguelay\corref{cor1}\fnref{APT}}
\ead{jade.giguelay@ens-paris-saclay.fr}

\author[maiage]{S.~Huet}
\ead{sylvie.huet@inra.fr}
\cortext[cor1]{Corresponding author}
\address[orsay]{Laboratoire de Math\'ematiques d'Orsay, Universit\'e Paris-Sud, CNRS,
  Universit\'e Paris-Saclay, 91405 Orsay, France}
\address[maiage]{MaIAGE INRA, Universit\'e Paris-Saclay, 78350 Jouy-en-Josas, France}
\fntext[APT]{Present address: AgroParisTech,  Paris 5e, France}

\begin{abstract}
We develop here several goodness-of-fit tests for testing the $k$-monotonicity of a discrete
density,  based on the empirical
distribution of the observations. Our tests are non-parametric, easy to implement and 
are proved to be asymptotically of the desired level and
consistent. We propose an estimator of the degree of $k$-monotonicity of the
distribution based on the non-parametric goodness-of-fit tests. We apply  our work
to the  estimation of the total  number of classes in a population. A
large simulation study allows to assess the performances of our procedures.
\end{abstract}

\maketitle

\section{Introduction}

The estimation of the distribution of  categorical  variables is an  important issues in statistical
research. 
For modeling count data parametric
models or nonparametric extensions such as mixtures of Poisson
distributions are very popular. 
 An alternative to these nonparametric modelings is to
consider a shape constraint on the underlying probability mass
function. Such approach  may be well
adapted in some situations because it  combines  the
straightforwardness of parametric  models (no choice of parameter is
left  to  the  user)   and  the  great  flexibility  of  nonparametric
estimation.  Moreover  shape
constraint  arises  naturally in  many  frameworks  such as
insurance~\cite{KACEM20151774}, reliability studies~\cite{Reboul2005}, epidemiology~\cite{balabdaoui2015maximum} or ecology~\cite{durot2015nonparametric, capture}.

Several authors have considered the problem of estimating a discrete
density under 
shape constraints. Balabdaoui et al.~\cite{balabdaoui2013asymptotics} considered the maximum-likelihood
estimator under constraint of log-concavity
and Balabdaoui and Jankowski~\cite{balabdaoui2015maximum} under constraint of unimodality. Jankowski and Wellner~\cite{jankowski2009estimation} studied the asymptotic
properties of several estimators of the density under assumption of
monotonicity. Durot et al.~\cite{durot_least} proposed a least-squares
estimator under convexity constraint while Giguelay~\cite{giguelay2017estimation}
considered  $k$-monotonicity constraint. The case $k=1$ corresponds to
monotonicity, the case $k=2$ to convexity, and the more $k$ increases, the
more the  density is hollow.

The constraint of $k$-monotonicity is especially 
suitable when one aims to estimate the unknown number of classes or
categories in a population. One of the main approaches to deal with
that problem consists in estimating the distribution of  the observed abundances for a series of
classes, from which the  estimation of the total number of  classes is
deduced. See Bunge and Fitzpatrick~\cite{bunge1993estimating} for a review of the
different approaches to deal with that problem. Durot et al.~\cite{durot2015nonparametric, capture} proposed
an estimator of the total number of classes based on an estimator of
the abundance distribution under the constraint of
convexity. Giguelay~\cite{theseJade} generalises their work to
$k$-monotonicity. 
Chee and Wang~\cite{chee2016nonparametric} proposed to model the
abundance distribution of species with a mixture of discrete beta
 distributions, such a mixture being $k$-monotone. These authors underlined that
 their model {\sl is
particularly suitable when a population is dominated by a large number
of rare species}.

In order to validate the chosen model before estimating the number of
classes, we propose a  
goodness-of-fit test for testing $k$-monotonicity.
 To the best of our knowledge, very few works are
available for testing a shape constraint on a discrete density: Akakpo
et al.~\cite{akakpo2014testing} proposed a procedure
for testing monotonicity ($k=1)$, while Durot et al.~\cite{capture}
and Balabdaoui et al.~\cite{balabdaoui2017testing} considered the problem of testing
convexity ($k=2$). The testing procedures they proposed rely on the asymptotic distribution of some
distance between the empirical distribution and the estimation of the
density under the shape constraint. This approach presents several
difficulties. It needs the calculation of the asymptotic distribution
of the test statistic under the null hypothesis which proves to be a
difficult problem even for $k=2$ both from a theoretical and a
computational  point of view.

We develop here several goodness-of-fit tests for $k$-monotonicity of a discrete
density,  based on the empirical
distribution of the observations. Our tests are non-parametric in the
sense that there is no parametric assumption on the underlying true
distribution of the observations.   The procedures
are easy to implement and 
are proved to be asymptotically of the desired level and
consistent. We carry out a large simulation study in order to assess
the performances of our procedures for finite sample size. From this
study, it appears
that the asymptotic specifications are achieved when the number of
observations is very large.  In order to
evaluate the intrinsic difficulty of these non-parametric procedures, 
we compare the efficiency of our
procedures to the one of parametric procedures constructed under the
assumption of  Poisson densities. This work is presented in
Section~\ref{Testk.st}. 

Next, in Section~\ref{Estk.st}, we propose an estimator of the degree of $k$-monotonicity of the
distribution based on the non-parametric goodness-of-fit tests. We
show that, if the true underlying distribution is $k$-monotone, then
the probability for our estimator $\widehat{k}$ to be  less than $k-1$
is smaller than  the chosen level of the  testing procedure. On the
other way, if the true underlying distribution is $k$-monotone but
not
$k+1$-monotone,  the probability for $\widehat{k}$ to be greater than
$k+1$ tends to zero. 

Finally, in Section~\ref{Estim.st}, we apply  our work
to the  estimation of the total  number of classes in a population ,
denoted $N$,
under the assumption that the abundances of the $N$ classes are i.i.d.
with common distribution $p = (p_0, p_1, \ldots)$ where for any
integer $j\leq 0$, $p_{j}$ is the
probability to observed a class $j$ times. Generalizing the work
of Durot et al.~\cite{durot2015nonparametric} we define a ``$k$-monotone abundance
distribution'' in order to make the total number of classes
identifiable. For each $k$, we are able to calculate an estimator of
$N$. At the same time, using the previous testing procedures,  we
estimate $k$, which  leads to a final estimator of $N$. This procedure is
illustrated in Section~\ref{realData.st} on three examples given in the litterature.

A small conclusion is given in Section~\ref{Conclusion-test-st} and all the
proofs are postponed to Section~\ref{proofs-tests.st}.

\section{Testing the $k$-monotonicity of a discrete distribution}
\label{Testk.st}

We
present   $k$-monotonicity testing 
procedures for any discrete distribution $p$ defined on a finite
support included in $\{0, \ldots, \tau \}$ for some unknown integer
$\tau$. Our results may be generalised to the case $\tau=\infty$ \\

Let us give the definition of $k$-monotonicity of a discrete
distribution. 

\begin{df}
Let $k\geq 1$ and for all $j\in\N$, let $\Delta^{k}p_j$ be the $k^{{\rm th}}$ differential operator of
$p$ defined as follows: 
\begin{eqnarray}
\Delta^{1}p_j & = & p_{j+1}-p_j \nonumber\\
\Delta^{k}p_j & = & \Delta^{k-1}p_{j+1}-\Delta^{k-1}p_j. \label{defDelta.eq}
\end{eqnarray}

A discrete distribution $p$ on $\N$ is $k$-monotone if and only if
\begin{equation*}
\nabla^{k}p_j = (-1)^{k} \Delta^{k}p_j  \geq 0, \mbox{for all } j \in \N.
\end{equation*}
\end{df}
It is easy to see that 
\begin{equation*}
 \nabla^{k}p_j = \sum_{h=0}^{k} (-1)^{h} C_{k}^{h} p_{j+h}.
\end{equation*}

It can be shown, see~\cite{giguelay2017estimation}, that if $p$ is $k$-monotone, then $p$ is strictly
$l$-monotone for all $1 \leq l \leq k-1$.
Moreover $p$ can be decomposed into a mixture of polynomial
distributions of order $k$~\cite{lefevre2013multiply}. More precisely, for all integer $j \in \N$
\begin{equation}
p_j  = \sum_{\ell \geq 0} \pi^{k}_{\ell} Q^{k}_{\ell}(j)
\label{QkDec.eq}
\end{equation}
where 
\begin{equation}
\pi^{k}_{\ell}  = C_{k+\ell}^{k}  \nabla^{k}p_\ell \mbox{ for all }
\ell \in \N,
\label{pi.eq}
\end{equation}
and where $Q^{k}_{\ell}$ is the $k$-monotone distribution defined as 
\begin{equation}
 Q^{k}_{\ell}(j) = \frac{C_{k-1+\ell-j}^{k-1}}{C_{k+\ell}^{k}}
 I(j\leq \ell),
\label{Qk.eq}
\end{equation}
where $I$ denotes the indicator function. 

The support of the distribution $\pi$ is the set of integers $j$ such
that $\nabla^{k} p_{j}$ is strictly positive. Such integers are
called the $k$-knots of $p$.

Let $X_1, \ldots, X_{d}$ be a $d$-sample with distribution $p$ and $f$
the relative frequencies: for all $j \geq 0$
\begin{eqnarray*}
 p_{j} & = & P( X_{i} =j) \\
f_j & = & \frac{1}{d}\sum_{i=1}^{d} I(X_{i} =j).
\end{eqnarray*}

We propose to test the null hypothesis that $p$ is $k$-monotone
considering the fact that  if $\nabla^{k}p_{j}$ is negative for some $j
\geq 0$, then $p$ is not $k$-monotone. Therefore we propose to reject
the $k$-monotonicity of $p$ if one of the estimators $\nabla^{k}f_{j}$
of $\nabla^{k}p_{j}$ is negative enough.

\subsection{Testing procedures and theoretical properties}
Let us begin with two testing procedures. The first one,
  denoted {\bf P1}, rejects the null
hypothesis if the minimum of the $\nabla^{k}f_{j}$'s is smaller than some
negative threshold, while the second one,
  denoted {\bf P2}, rejects the null
hypothesis if one of the hypothesis ``$\nabla^{k}p_{j} \geq 0$'' is
rejected. Procedure  {\bf P2} is a standardized version
of Procedure {\bf P1}.

Let us introduce the following notations:
\begin{itemize}
\item $\Gamma$ is  the matrix
     with components $\Gamma_{j j'} = - p_{j} p_{j'} $
    if $j\neq j'$ and $\Gamma_{j j} =
    p_{j}(1-p_{j})$ for $0 \leq j, j' \leq \tau$, and $\Gamma^{1/2}$
     its square-root such that $\Gamma^{1/2}\Gamma^{1/2}=\Gamma$. 
\item $A^{k}$ is the matrix whose lines $A^{kT}_{j}$ satisfy 
    $\nabla^{k}p_{j} = A^{k T}_{j} p$ for $j=0, \ldots, \tau-1$.
\item $M^{k}$ is the square-root of the matrix 
    $A^{k}\Gamma A^{kT}$: $M^{k}M^{k} =
    A^{k}\Gamma A^{k T}$
\item $(\Z_{j'}, j'=0, \ldots \tau-1)$ are i.i.d. $\cN(0,1)$
  variates, and $\Z$ is the random vector with components
    $\Z_{j'}, j'=0, \ldots \tau-1$.
\item For $0 < \alpha < 0.5$,  
\begin{equation}
q^{k}_{\alpha} = \inf_{q} \left\{ 
\P\left(\min_{0\leq j\leq \tau-1} \sum_{j'=0}^{\tau-1} M^{k}_{j
 j'} \Z_{j'} \leq q\right) = \alpha  
\right\},
\label{AsThatk.eq}
\end{equation} 
\begin{equation*}
 \label{ualpha.eq}
u^{k}_{\alpha} = \max_{0\leq u \leq 1} \left\{\P \left( \min_{0\leq j
      \leq \tau-1} 
\left\{ A_{j}^{k T} \Gamma^{1/2} \Z  -
  \nu_{u} \sqrt{ A^{k T}_{j}\Gamma A^{k}_{j}}
\right\} \leq 0 \right) = \alpha\right\},
\end{equation*}
where  $\nu_{u}$ is the $u$-quantile
of a $\cN(0,1)$ variable.
\item $\widehat{\tau}$ the maximum of the
support of the empirical 
distribution, $\widehat{\tau} =
    \max_{i=1, \ldots D} X_{i}$
\item $\widehat{\Gamma}$, $\widehat{M}^{k}$, $\widehat{\Z}$,
  $\widehat{q}^{k}_{\alpha}$, $\widehat{u}^{k}_{\alpha}$
   are defined
  as above with $f$ instead of $p$ and $\widehat{\tau}$ instead of
  $\tau$.
\end{itemize}

\paragraph{Testing procedures}

\begin{description}
\item[P1]  The rejection region for testing that $p$ is
    $k$-monotone is defined as 
\begin{equation*}
\left\{\widehat{\T}^{k} \leq \widehat{q}^{k}_{\alpha}\right\} \mbox{
  where } \widehat{\T}^{k} = \sqrt{d} \min_{0\leq j \leq \widehat{\tau}-1}
     \nabla^{k}{f}_{j}.
\end{equation*}
Let us note that the threshold $\widehat{q}^{k}_{\alpha}$ defined above, is the $\alpha$-quantile of the
    conditional  distribution  given $X_{1}, \ldots, X_{d}$ of 
\begin{equation}
 \widehat{\U}^{k} = \min_{0\leq j\leq \widehat{\tau}-1} \sum_{j'=0}^{\widehat{\tau}-1} \widehat{M}^{k}_{j
 j'} \widehat{\Z}_{j'}.
\label{Uchapk.eq}
\end{equation}
It is calculated  by simulation.

\item[P2] The second procedure will reject the null hypothesis if
   the 
minimum of  $\nabla^{k}f_{j}$ minus some threshold depending on $j$ is
negative. Precisely  the rejection region for testing that $p$ is $k$-monotone is
defined as 
\begin{equation*}
 \left\{\widehat{\cS}_{\alpha}^{k} \leq 0\right\} \mbox{ where }
 \widehat{\cS}^{k}_\alpha = \min_{0\leq j \leq \widehat{\tau}-1}
 \left\{\sqrt{d} \nabla^{k}{f}_{j} - \nu_{\widehat{u}^{k}_{\alpha}} \sqrt{ A^{k T}_{j}\widehat{\Gamma} A^{k}_{j}}\right\}.
\end{equation*}
The quantity $\widehat{u}^{k}_{\alpha}$ is calculated by simulation.
\end{description}

We also propose a bootstrap procedure for calculating either the quantiles
$\widehat{q}^{k}_{\alpha}$ or the $\nu_{u}$ for $u$ in a grid of
values. These procedures called {\bf P1boot} and {\bf P2boot} are
described in Section~\ref{boot.st}

The two following theorems give the asymptotic properties of the
     testing procedures. Their proof are given in Section~\ref{proofs-tests.st}.

\begin{theo}
\label{TestkLevel.th}
{\bf Level of the test.} 
~
\\
\underline{Let $p$ be a $k$-monotone distribution} with finite support. The
testing procedures have asymtotic level $\alpha$:
\begin{equation*}
 \lim_{d \rightarrow \infty }\P\left( \widehat{\T}^{k} \leq
  \widehat{q}^{k}_{\alpha} \right)   \leq  \; \alpha 
,\;\lim_{d \rightarrow \infty }
\P\left( \widehat{\cS}^{k}_{\alpha} \leq 0 \right)   \leq  \;\alpha 
\end{equation*}
\underline{If $p$ is a strictly $k$-monotone distribution} with finite support,
then we have the following result
\begin{description}
\item[P1] Let $\sigma^{k} = \max_{0 \leq j \leq \tau-1}
 \sqrt{ \sum_{j'=0}^{\tau-1}(M^{k}_{jj'})^{2} }$ and $\beta>0$.
If the distribution $p$ satisfies the following property
\begin{equation*}
\min_{0 \leq j \leq\tau-1} \nabla^{k} p_{j} \geq \sqrt{\frac{2}{d}} 
\sigma^{k} \sqrt{\log \frac{\tau}{\beta}},
\end{equation*} then  
\begin{equation*}
\lim_{d \rightarrow \infty }\P\left( \widehat{\T}^{k} \leq
  \widehat{q}^{k}_{\alpha} \right)  \leq \beta.
\end{equation*}
\item[P2] Let $\zeta_{j}^{k} = \sqrt{A_{j}^{k T} \Gamma  A_{j}^{k}}$,
  and $ \zeta^{k} = \max_{j}\zeta_{j}^{k}$, and let $\beta>0$.
If the distribution $p$ satisfies the following property 
\begin{equation*}
\min_{0 \leq j \leq\tau} \nabla^{k} p_{j} \geq \sqrt{\frac{2}{d}} 
\zeta^{k} \sqrt{\log\frac{\tau}{\beta}},
\end{equation*}
  then 
\begin{equation*}
 \lim_{d \rightarrow \infty }
\P\left( \widehat{\cS}^{k}_{\alpha} \leq 0\right)  \leq \beta.
\end{equation*}
\end{description}
\end{theo}

In particular, if the
distribution $p$ is strictly $k$-monotone and satisfies the above
condition for $\beta$ that tends to 0, for example
$\beta = 1/\sqrt{d}$, then the level of the test tends to 0.

\begin{theo}
\label{TestkPower.th}
{\bf Power of the test.}\\ \underline{Let $p$ be a $k$-monotone distribution, but not a $(k+1)$-monotone
distribution} and $\beta>0$. 
\begin{description}
\item[P1] Let $\sigma^{k} = \max_{0 \leq j \leq \tau-1}
 \sqrt{ \sum_{j'=0}^{\tau-1}(M^{k}_{jj'})^{2} }$. If $p$ satisfies the following
condition: 
\begin{equation*}
 \exists j_{0},    \nabla^{k+1}p_{j_{0}} + \frac{1}{\sqrt{d}}  
\left(\sigma^{k+1} \sqrt{2\log\frac{\tau}{\alpha}}
 +\zeta^{k+1}_{j_{0}} \sqrt{-2 \log \beta}\right) \leq 0,
\end{equation*}
then 
 we have the following result:
\begin{equation*}
\lim_{d  \rightarrow \infty} 
\P \left(  \widehat{\T}^{k+1} \geq
  \widehat{q}^{k+1}_{\alpha} \right) \leq \beta.
\end{equation*}
\item[P2] Let $\zeta_{j}^{k} = \sqrt{A_{j}^{k T} \Gamma  A_{j}^{k}}$. If $p$ satisfies the following
condition: 
\begin{equation*}
\exists j_0,\; \nabla^{k+1}p_{j_{0}} + \frac{1}{\sqrt{d}}  
\left(
\sqrt{2 \log\frac{\tau}{\alpha}} 
 +  \sqrt{-2\log \beta}\right) \zeta^{k+1}_{j_{0}} \leq 0, 
\end{equation*}
then  we have the following result:
\begin{equation*}
\lim_{d  \rightarrow \infty} 
\P \left(  \widehat{\cS}_{\alpha}^{k} \geq 0 \right) \leq \beta.
\end{equation*}
\end{description}
\end{theo}

In order to evaluate the performances of the two testing procedures
for finite sample size, 
we carry out a simulation study. In Section~\ref{SimPoisson.st} we
consider Poisson distributions with parameters chosen to ensure
$k$-monotonicity, and in Section~\ref{SimSplines.st} we consider
mixtures of Splines distributions.

\subsection{\label{SimPoisson.st}Simulation for Poisson  distributions}

\subsubsection{\label{PoissonDist.st}Poisson distribution and $k$-monotonicity}

We carry out a simulation study, considering empirical
distributions simulated according to Poisson distributions with
parameters $\lambda^{h}$ chosen as follows: for all  $\lambda \leq \lambda^{h}$ then $p \sim
\cP(\lambda)$ is at least $h$-monotone, and for all  $\lambda \in \left] \lambda^{h}, \lambda^{h-1} \right]$, then  $p \sim
\cP(\lambda)$ is $(h-1)$-monotone but not $h$-monotone. For $h \in \{1,
\ldots, 10\}$ the values of $\lambda^{h}$, calculated numerically, are given at
Table~\ref{lambdaValues.tab}. Note that by choosing these values of
$\lambda$ for our simulation study, we are in the best scenario to reject
$H_{k}$ when $k=h+1$, since the Poisson distribution with parameter
$\lambda^{h}$ is the most distant from the set of $k$-monotone Poisson distributions.

\FloatBarrier
\begin{table}[!h]
\begin{center}
\begin{tabular}{r|ccccccccccc}
$h$ & 0 &   1 & 2&  3 & 4 & 5 & 6 & 7 & 8 & 9 & 10 \\
$\lambda^{h}$ & 2 & 1 & 0.5857 & 0.4157&  0.3225&  0.2635 & 0.2228 & 0.193 & 0.1703 & 0.1523 &  0.1377  \\
\end{tabular}
\end{center}
\caption{\label{lambdaValues.tab} For $h \geq 0$, values of
  $\lambda^{h}$ used in the simulation study: the data are generated
  according to the distributions $\cP(\lambda^{h})$.}
\end{table}
\FloatBarrier

\begin{figure}
\begin{center}
\includegraphics[height=.45\textwidth, width=.90\textwidth,angle=0]{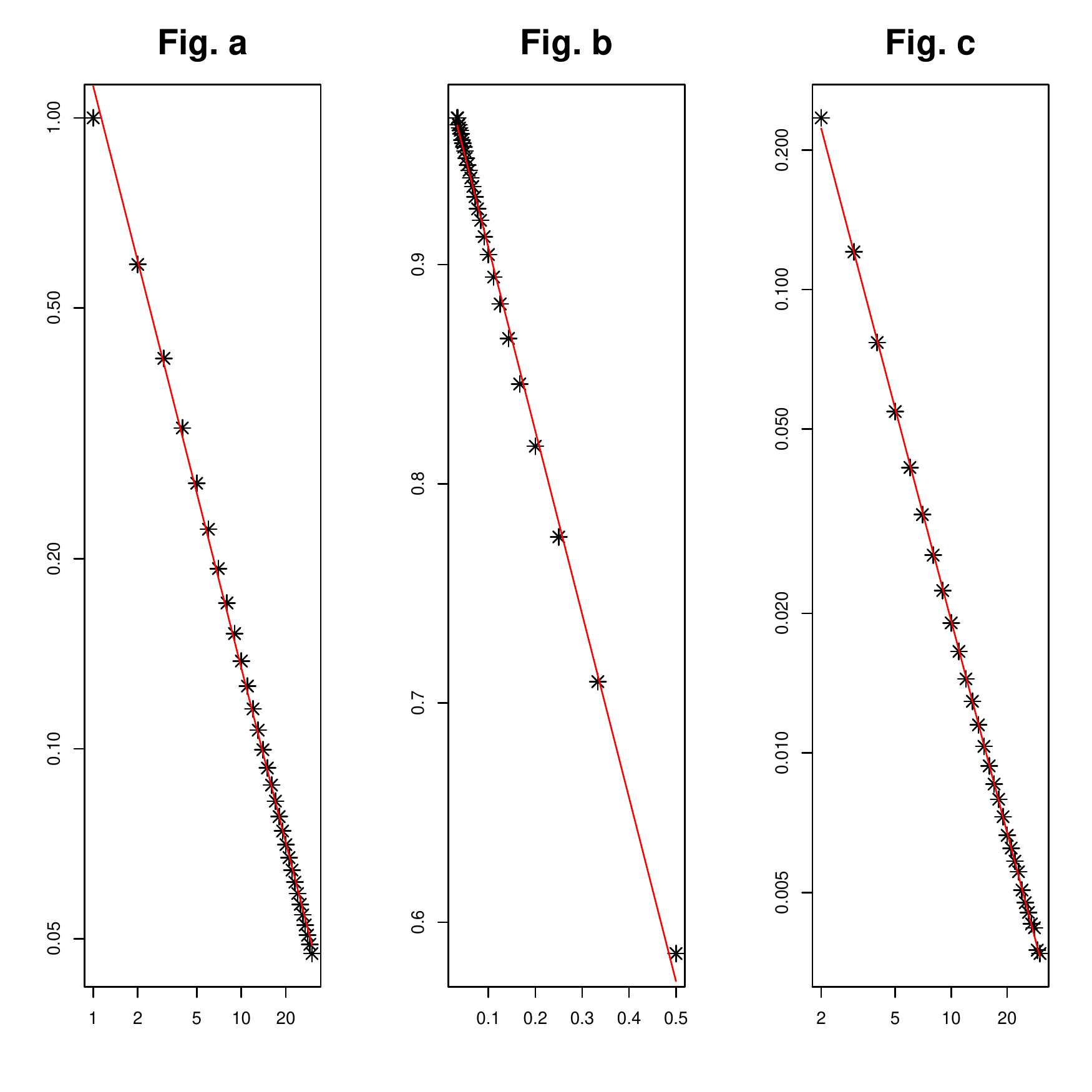}
\end{center}
\caption{\label{lambdak.fig} Variation of $\lambda^{h}$ versus $h$ and
  fitted line in red. \textbf{Fig. a}:  $\log\left(\lambda^{h}\right)$ versus $\log(h)$,
  fitted $=0.12-0.92 \log(h)$.
\textbf{Fig. b}: $\lambda^{h+1}/\lambda^{h}$ versus $1/(h+1)$,
  fitted $=0.99-0.84/(h+1)$.
\textbf{Fig. c}: $\log\left(\sqrt{\lambda^{h}} -
  \sqrt{\lambda^{h+1}}\right)$ versus $\log(h+1)$, fitted $=-0.45-1.52\log(h+1)$.}
\end{figure}

When $\lambda > 1$, the Poisson distribution $\cP(\lambda)$ is
unimodal. In the simulation study, we choose $\lambda^{0}=2$, to
represent non monotone distributions. For $h \geq 1$, the values of $\lambda^{h}$ and the differences
$\lambda^{h}-\lambda^{h+1}$ decrease with $h$, see  Figure~\ref{lambdak.fig}. Some numerical
calculations show that  $\lambda^{h}$ decreases approximatively as
$1.13/h^{0.9}$, $\lambda^{h+1}/\lambda^{h}$ decreases as
$1-0.84/(h+1)$, while $\sqrt{\lambda^{h}}-\sqrt{\lambda^{h+1}}$ decreases as
$0.64/(h+1)^{1.52}$ (this last result will be usefull later
on). This suggests that testing $H^{h+1}$ when $p \sim \cP(\lambda^{h})$
will be difficult for large $h$.
\\

\subsubsection{Simulation study}

\paragraph{Procedure {\bf P1}}
For each value of $h$, we estimate the rejection probabilities of
hypotheses $H^{k}$, for $k \in \{1, \ldots, 9\}$ on the basis of
$500$ runs. The results for procedure {\bf P1} are given in
Table~\ref{SimulPbRejetHkPoisson.tab}. 
\\

\begin{table}[htb]
\begin{center}
For $d=1000$, $p \sim \cP(\lambda^{h})$

\medskip

\begin{small}
\begin{tabular}{r|ccccccccccc}
    &  $h= 10$& $h= 9$ &$h= 8$&$h= 7$ &$h= 6$ &$h= 5$ &$h= 4$ &$h= 3$&$h=2$&$h= 1$& $h= 0$ \\\hline 
$k= 1$ & 0 & 0 & 0 & 0 & 0 & 0 & 0 & 0 & 0 & 0.036 & {\bf 1} \\
$k= 2$ & 0 & 0 & 0 & 0 & 0 & 0 & 0 & 0
& 0.058 & {\bf 0.990} &{\it 1} \\
$k= 3$ & 0 & 0 & 0 & 0 & 0 & 0 & 0 & 0.060
& {\bf 0.760} & {\it 0.958} & 0.370 \\
$k= 4$ & 0 & 0 & 0 & 0 & 0 & 0.002 & 0.030 & {\bf
  0.422} & {\it 0.814} & {\it 0.646} & 0.074 \\
$k= 5$ & 0 & 0 & 0 & 0 & 0 & 0.064 & {\bf 0.236} &
{\it 0.560} & {\it 0.678} & 0.268 & 0.042 \\
$k= 6$ & 0 & 0 & 0 & 0.010 & 0.060 & {\bf 0.216} & 0.378 & {\it 0.524} & 0.470 & 0.108 & 0.042 \\
$k= 7$ & 0 & 0 & 0.014 & 0.062 & {\bf 0.148} & 0.300 & 0.362 & 0.410 & 0.272 & 0.056 & 0.058 \\
$k= 8$ & 0.010 & 0.018 & 0.050 & {\bf 0.126} & 0.224 & 0.326 & 0.320 & 0.256 & 0.160 & 0.042 & 0.062 \\
$k= 9$ & 0.030 & 0.044 & {\bf 0.102} & 0.178 & 0.244 & 0.296 & 0.252 & 0.206 & 0.112 & 0.036 & 0.058 \\
$k= 10$& 0.080 & {\bf 0.080} & 0.142 & 0.206 & 0.240 & 0.268 & 0.192 & 0.158 & 0.090 & 0.042 & 0.060 \\
\end{tabular}
\end{small}
\end{center}

\begin{center}
For $d=5000$, $p \sim \cP(\lambda^{h})$
\begin{small}
\begin{tabular}{r|ccccccccccc}
&$h= 10$&$h= 9$&$h= 8$&$h= 7$&$h= 6$&$h= 5$&$h= 4$&$h=
3$&$h=2$&$h=1$&$h= 0$ \\ \hline
$k= 1$& 0 & 0 & 0 & 0 & 0 & 0 & 0 & 0 & 0 & 0.034 & {\bf 1}  \\
$k= 2 $& 0 & 0 & 0 & 0 & 0 & 0 & 0 & 0 & 0.056 & {\bf 1} & {\it 1}  \\
$k= 3 $& 0 & 0 & 0 & 0 & 0 & 0 & 0 & 0.062 & {\bf 1} & {\it 1} & {\it 0.992}  \\
$k= 4 $& 0 & 0 & 0 & 0 & 0 & 0 & 0.052 & {\bf 0.960} & {\it 1} & {\it 1} & 0.198  \\
$k= 5$& 0 & 0 & 0 & 0 & 0 & 0.062 & {\bf 0.766} & {\it 0.994} & {\it 1} & {\it 0.762} & 0.036  \\
$k= 6 $& 0 & 0 & 0 & 0 & 0.040 & {\bf 0.494} & {\it 0.904} & {\it 0.986} & {\it 0.962} & 0.260 & 0.018  \\
$k= 7 $& 0 & 0 & 0.002 & 0.056 & {\bf 0.370} & {\it 0.748} & {\it 0.910} & {\it 0.948} & {\it 0.694} & 0.092 & 0.036  \\
$k= 8 $& 0 & 0.004 & 0.060 & {\bf 0.256} & 0.580 & {\it 0.776} & {\it 0.852} & {\it 0.788} & 0.360 & 0.044 & 0.050  \\
$k= 9 $& 0.006 & 0.070 & {\bf 0.156} & 0.434 & {\it 0.650} & {\it 0.738} & {\it 0.746} & {\it 0.564} & 0.166 & 0.040 & 0.050  \\
$k=10$& 0.042 & {\bf 0.170} & 0.306 & 0.490 & {\it 0.628} & {\it 0.636} & {\it 0.584} & 0.368 & 0.090 & 0.042 & 0.048  \\
\end{tabular}
\end{small}
\end{center}
\caption{\label{SimulPbRejetHkPoisson.tab} 
  Procedure {\bf P1}: Estimated probabilities of  rejecting  the hypothesis $H^{k}$, for Poisson
  distributions with parameters $\lambda^{h}$ given at
  Table~\ref{lambdaValues.tab}. In bold character, the probablities of rejecting $H^{k}$ with $k=h+1$, for $h
  \geq 0$. In italic character, the probablities of  rejecting greater
  than $0.5$.
}
\end{table}

\begin{table}[htb]
\begin{center}
\begin{small}
\begin{tabular}{r|cccccccc}
 ~&$k= 3$&$k= 4$ &$k= 5$ &$k= 6 $ &$ k= 7 $ &$ k= 8 $ &$ k= 9 $ &$k=
                                                                  10 $
  \\ \hline
$\sqrt{d} \nabla^{k} p_0$ &
 -4.86& -7.79& -9.23&  -9.52&  -8.98 & -7.87&  -6.41 & -4.79 \\
$\sqrt{A^{k T}_{0} \Gamma A^{k}_{0}}$ & 2.08 & 3.07 & 4.48&   6.46 &  9.21 & 13.0 & 18.2&  25.2 \\
$q^{k}_{\alpha}$ & -3.39& -5.22& -7.34& -10.7& -15.3& -22.0& -29.9& -41.3 \\
\end{tabular}
\end{small}
\end{center}
\caption{\label{suppl.tb} 
For the Poisson
  distribution with parameters $\lambda^{2}=0.5857$, for
  $d=1000$, values 
  of $\sqrt{d} \min_{j \geq 0}\left\{ \nabla^{k} p_j\right\} =
  \sqrt{d} \nabla^{k} p_0$, of its standard error $\sqrt{A^{k T}_{0}
    \Gamma A^{k}_{0}}$ and of $q^{k}_{\alpha}$ for $\alpha=5\%$.
}
\end{table}

It appears that the level of the test based on procedure {\bf P1} is close to $\alpha$ when $k=h$ and
equals 0 as soon as $k$ is smaller than $h+1$. This result confirms
Theorem~\ref{TestkLevel.th}  that states that the level of the test of
the hypothesis $H^{k}$ tends to 0 if the distribution is strictly
$k$-monotone.
\\

As expected, the power of the test of the hypothesis $H^{k+1}$ when
$h=k$ decreases with $k$. 
Moreover, for a fixed value of $h$, and for $k \geq h+1$,  the power of the test
of the hypothesis $H^{k}$ first increases with $k$, then decreases
with $k$.  In fact, the decreasing of the power for large values of
$k$ may be explained as follows: when $k$ increases, the
variances of the components of 
$\widehat{\T}^{k}$ increase, and the $5\%$-quantiles of the variate
$\widehat{\U}^{k}$ 
given at Equation~\eref{Uchapk.eq} becomes strongly negative, see
Table~\ref{suppl.tb}.  
\\

This simulation leads to the following remarks:
\begin{rmk}
It confirms that the procedure {\bf P1} for testing  $H^{k}$, when
the true distribution is $(k-1)$-monotone, lacks of
power when $k$ is large. For example if the true distribution is
$4$-monotone, the hypothesis $H^{5}$ will not be rejected with
probability greater than $0.76$ (respectively 0.23) if $d=1000$
(respectively $d=5000$). 
\end{rmk}
\begin{rmk}
It shows that the power of the test of hypothesis $H^{k}$ when
the true distribution is $h$-monotone, can be small when $k-h$ is
large. For example if the true distribution is convex ($h=2$), the
hypothesis  $H^{7}$ will not be rejected with
probability greater than $0.72$ if $d=1000$. 
Nevertheless, let us note that the power for testing $H^{3}$ is large
(it equals
$0.76$ for $d=1000$).
\end{rmk}

Finally, let us note that testing the hypothesis $H^{k+j}$ for $j \geq 1$, when
we rejected  $H^{k}$ is without interest, because we know that a
$(k+j)$-monotone distribution is necessarily $k$-monotone. Therefore a
natural idea is to modify the procedure
 in order to test the 
hypothesis $H^{k}$ if $H^{k-1}$ is not rejected.  In other words, if
$H^{k-1}$ is rejected, we decide that $H^{\ell}$ is rejected for all
$\ell \geq k$. The probabilities of not rejecting the hypotheses
$H^{k}$ are estimated  on the basis of
$500$ runs and reported in Table~\ref{SimulTestPoisson.tab}.

\begin{table}[htb]
\begin{center}
For $d=1000$, $p \sim \cP(\lambda^{h})$
\begin{small}
\begin{tabular}{r|ccccccccccc}
      &$h= 10$  &$h= 9$  &$h= 8$  &$h= 7$  &$h= 6$  &$h= 5$  &$h= 4$
      &$h= 3$  &$h= 2$  &$h= 1$ &$h= 0$ \\
$k= 1 $  &0 &0 &0 &0 &0 &0 &0 &0 &0 &0.036    &1 \\
$k= 2 $  &0 &0 &0 &0 &0 &0 &0 &0 &0.058 &{\bf 0.990 }   &1 \\
$k= 3 $  &0 &0 &0 &0 &0 &0 &0 &0.060 &{\bf 0.760} &0.994    &1 \\
$k= 4 $  &0 &0 &0 &0 &0 &0.002 & 0.030 &{\bf 0.422} &0.826 &0.994    &1 \\
$k= 5 $  &0 &0 &0 &0 &0 & 0.064 &{\bf 0.236} &0.560 &0.826 &0.994    &1 \\
$k= 6 $  &0 &0 &0 &0.010 & 0.060 &{\bf 0.216} &0.378 &0.566 &0.826 &0.994    &1 \\
$k= 7 $  &0 &0 &0.014 & 0.062 &{\bf 0.148} &0.300 &0.396 &0.568 &0.826 &0.994    &1 \\
$k= 8 $  &0.010 &0.018 &0.050 &{\bf 0.126}&0.224 &0.330 &0.398 &0.568 &0.826 &0.994    &1 \\
$k= 9 $  &0.030 &0.044 &{\bf 0.102} &0.180 &0.248 &0.338 &0.398 &0.568 &0.826 &0.994    &1 \\
$k= 10$  &0.080 &{\bf 0.080} &0.142 &0.212 &0.258 &0.346 &0.398 &0.568 &0.826 &0.994    &1  \\
\end{tabular}
\end{small}
\end{center}

\begin{center}
For $d=5000$, $p \sim \cP(\lambda^{h})$
\begin{small}
\begin{tabular}{r|ccccccccccc}
      &$h= 10$  &$h= 9$  &$h= 8$  &$h= 7$  &$h= 6$  &$h= 5$  &$h= 4$  &$h= 3$  &$h= 2$  &$h= 1$ &$h= 0$ \\
$k= 1$  &0 &0 &0 &0 &0 &0 &0 &0 &0 &0.034    &1 \\
$k= 2$  &0 &0 &0 &0 &0 &0 &0 &0 &0.056 &{\bf 1}    &1 \\
$k= 3$  &0 &0 &0 &0 &0 &0 &0 &0.062 &{\bf 1} &1    &1 \\
$k= 4$  &0 &0 &0 &0 &0 &0 &0.052 &{\bf 0.960} &1 &1    &1 \\
$k= 5$  &0 &0 &0 &0 &0 &0.062 &{\bf 0.766} &0.994 &1 &1    &1 \\
$k= 6$  &0 &0 &0 &0 &0.040 &{\bf 0.494} &0.904 &0.994 &1 &1    &1 \\
$k= 7$  &0 &0 &0.002 &0.056 &{\bf 0.370} &0.748 &0.918 &0.994 &1 &1    &1 \\
$k= 8$  &0  &0.004 &0.060 &{\bf 0.256} &0.580 &0.780 &0.922 &0.994 &1 &1    &1 \\
$k= 9$  &0.006 &0.070 &{\bf 0.156} &0.434 &0.652 &0.782 &0.922 &0.994 &1 &1    &1 \\
$k= 10$ &0.042 &{\bf 0.170} &0.306 &0.494 &0.664 &0.782 &0.922 &0.994 &1 &1    &1 \\
\end{tabular}
\end{small}
\end{center}
\caption{\label{SimulTestPoisson.tab} 
  Procedure {\bf P1}: Estimated probabilities of  rejecting  the
  hypothesis $H^{k}$ knowing that  $H_{k-1}$ is not rejected, for Poisson
  distributions with parameters $\lambda^{h}$ given at
  Table~\ref{lambdaValues.tab}.
In bold character, the probablities of rejecting $H^{k}$ with $k=h+1$, for $h
  \geq 0$. 
}
\end{table}

\paragraph{Comparison with procedures {\bf P2} }

The results using  procedures {\bf P2} are slightly worse
 or equivalent to  those of procedure {\bf P1}, see Table~\ref{PoissonP2P3.tb}. 
This is easily
 understandable in the case of Poisson distribution 
 the rejection of the null hypothesis
 lies essentially on $\nabla^{k} p^{h}_{0}$, whatever the procedure.

\begin{table}
\begin{center}
For $d=1000$, $p \sim \cP(\lambda^{h})$

\begin{small}
\begin{tabular}{rccccccccccc}
& \multicolumn{11}{c}{Procedure {\bf P2}} \\ \hline
       &$h=10$&$h=9$&$h=8$&$h=7$&$h=6$&$h=5$&$h=4$&$h=3$&$h=2$&$h=1$&$h=0$\\
$k=h$  &0.024 &0.024&0.016&0.022&0.016&0.016&0.010&0.024&0.014&0.006&\\ 
$k=h+1$&      &0.044&0.036&0.064&0.072&0.094&0.126&0.228&0.518&0.956&1\\\\
\end{tabular}
\end{small}

For $d=5000$, $p \sim \cP(\lambda^{h})$

\begin{small}
\begin{tabular}{rccccccccccc}
& \multicolumn{11}{c}{Procedure {\bf P2}} \\ \hline
       &$h=10$&$h=9$&$h=8$&$h=7$&$h=6$&$h=5$&$h=4$&$h=3$&$h=2$&$h=1$&$h=0$\\
$k=h$  &0.032 &0.022&0.018&0.014&0.004&0.016&0.008&0.008&0.016&0.010& \\ 
$k=h+1$&      &0.086&0.094&0.136&0.180&0.312&0.540&0.860&1&1&1\\
\end{tabular}
\end{small}
\end{center}
\caption{\label{PoissonP2P3.tb} Procedures {\bf P2} : Estimated
 probabilities of rejecting $H^{h}$ and $H^{h+1}$ for Poisson distribution with
 parameters $\lambda^{h}$.}
\end{table}

\subsubsection{Comparison with parametric testing procedures}

Our simulation study showed that the testing procedure lacks of power both
when $k$ and $h$ increase. We would like to understand if
this difficulty is inherent to the testing problem, or comes from a
bad choice of the testing procedure. For the sake of simplicity we
focus on the power when testing $H^{k}$ with $k=h+1$.
\\

To answer our
question, we will consider a parametric framework where the distribution is known to be a
Poisson distribution. It is then  possible to propose a
parametric testing procedure for testing
the $k$-monotonicity, where the null hypothesis is a simple
hypothesis. 
This parametric
framework will 
constitute a kind of benchmark for the performances of the test. 
\\

We propose the following parametric testing procedure: for
$h \geq 1$, we test the null hypothesis
that $p$ is at least $(h+1)$-monotone against the alternative that $p$
is $h$-monotone but not $(h+1)$-monotone. 
In other words, we test
\begin{equation*}
  p \sim \cP(\lambda^{h+1}) \mbox{ against }   p \sim
 \cP(\lambda) \mbox{ with } \lambda \in \left] \lambda^{h+1},
 \lambda^{h} \right],
\end{equation*}
assuming that $X_{1}, \ldots, X_{d}$ are i.i.d. with distribution
$\cP(\lambda)$. 
\\

For this testing procedures, as well as for procedure {\bf P1} (see Theorem~\ref{TestkPower.th}), the rate of testing is the parametric rate
$1/\sqrt{d}$. Nevertheless the power depends also strongly on
$k$. Instead of studying the decreasing of the power versus $k$, that
depends also on $d$,  we compare the efficiencies of the procedures by 
calculating the minimal number of observations such that the power of
the test is greater that some fixed value, and study how this number
increases with $k$. Let us describe how these quantities are
calculated according to the testing procedure.

We denote  $p=p^{h}$ the density of a Poisson distribution with parameter
$\lambda^{h}$. 

\paragraph{Efficiency for the procedure P1} Let $q_{\alpha}^{h+1}$ be defined at
Equation~\eref{AsThatk.eq} calculated for $p=p^{h+1}$ and $\tau$ chosen 
large enough to get $\sum_{j=0}^{\tau} p_{j}
\approx 1$. 

Let $M^{h+1, h}$ be the square-root of the matrix $A^{h+1} \Gamma^{h}
(A^{h+1})^{T}$, where $\Gamma^{h}$ is calculated for $p=p^{h}$. 
\\

For a sample size $d$,
let $\pi_{\alpha, d}^{h}$ be defined as follows:
\begin{equation*}
 \pi_{\alpha, d}^{h} = \P\left( \min_{0 \leq j \leq \tau-1} 
\left\{ \sqrt{d} (A_{j}^{h+1})^{T} p^{h} + \sum_{j'=0}^{\tau-1} M_{j
    j'}^{h+1, h} \Z_{j'}\right\} \leq q_{\alpha}^{h+1}
\right).
\end{equation*}

Following the proof of Theorem~\ref{TestkPower.th}, it is easy to show
that when $d$ is large enough, $\pi_{\alpha, d}^{h}$  approximates the
power of the test of the hypothesis $H^{k}$ in $\lambda=\lambda^{h}$, with $k=h+1$:
\begin{equation*}
 \P_{p=p^{h}}\left( \widehat{\T}^{h+1} \leq \widehat{q}_{\alpha}^{h+1}\right) =
 \pi_{\alpha, d}^{h} + o_{p}(1).
\end{equation*}

Let $\beta>0$, for each $h$,  we
determine the value of $d$ for which the power of the test is greater
than $1-\beta$:
\begin{equation*}
d_{{\rm P1}} ^{h} = \inf_{d} \left\{ \pi_{\alpha, d}^{h} \geq 1-\beta\right\}.
\end{equation*}

The values of $ \pi_{\alpha, d}^{h}$ and $d_{{\rm P1}} ^{h}$ are
calculated by simulation.

\paragraph{Efficiency for the parametric procedure} 

The parametric testing procedure is based on $\bar{X}$, the mean of
the observations. If $p=p^{h+1}$, $d \bar{X}$ is distributed as a
Poisson variable with parameter $d \lambda^{h+1}$. In what follows,
this distribution will be approximated by a Gaussian distribution with
mean and variance equal to $d \lambda^{h+1}$. 
\\

The null hypothesis will be rejected for large values of
$\bar{X}$. More precisely, under the Gaussian approximation, 
we get the following results:
\begin{eqnarray*}
 \P_{p=p^{h+1}}\left( \bar{X} > \lambda^{h+1} + 
\sqrt{\frac{\lambda^{h+1}}{d}} \nu_{1-\alpha}
\right) & = & \alpha \\
\P_{p=p^{h}}\left( \bar{X} > \lambda^{h+1} + 
\sqrt{\frac{\lambda^{h+1}}{d}} \nu_{1-\alpha}
\right) & = & 1 - \Phi\left(
\sqrt{\frac{d}{\lambda^{h}}} \left(\lambda^{h+1} - \lambda^{h}\right)
+ \sqrt{\frac{\lambda^{h+1}}{\lambda^{h}}} \nu_{1-\alpha}
\right)
\end{eqnarray*}
and
\begin{equation}
 d_{{\rm P}} ^{h} = \left(
\frac{\sqrt{\lambda^{h}} \nu_{\beta} - \sqrt{\lambda^{h+1}}
  \nu_{1-\alpha}}{
\lambda^{h} - \lambda^{h+1}}
\right)^{2}.
\label{dP.eq}
\end{equation}

\paragraph{Comparison of the two procedures}

Taking $\beta=\alpha$, we compare $d_{{\rm P}}^{h}$ and $d_{{\rm
    P1}}^{h}$.

\begin{figure}
\begin{center}
\includegraphics[height=.45\textwidth, width=.80\textwidth,angle=0]{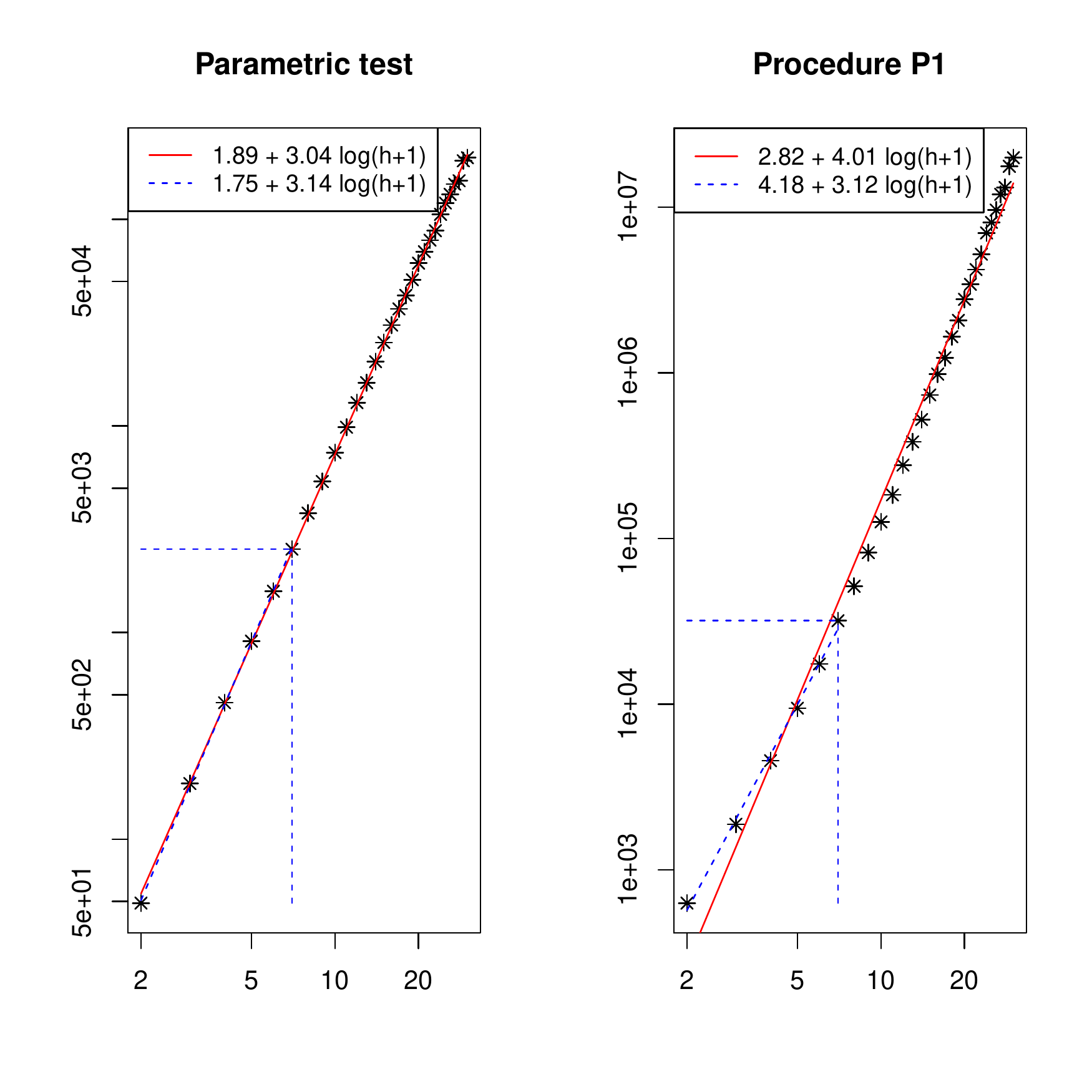}
\end{center}
\caption{\label{Compar.n.fig} Variation of $d_{{\rm P}}^{h}$ (left
  hand side) and
  $d_{{\rm P1}}^{h}$ (right hand side) versus $h+1$, in $\log$-scale. The red line (respectively the
  dashed blue line) corresponds to a
  linear fit in the log-scale for $h \in {1, \ldots, 29}$
  (respectively $h \in {1, \ldots, 6}$).}
\end{figure}

For 
    the parametric test we get that $d_{{\rm P}}^{h}$ is of order
    $(h+1)^{3}$, see Figure~\ref{Compar.n.fig}. This corresponds to the order of magnitude given by
    Equation~\eref{dP.eq}. Indeed when $\alpha=\beta$
\begin{equation*}
  d_{{\rm P}} ^{h} 
=
\frac{\nu_{1-\alpha}^{2}}{
\left(\sqrt{\lambda^{h}} - \sqrt{\lambda^{h+1}}\right)^{2}},
\end{equation*}
which varies as $(h+1)^{3}$ as it was shown in Figure~\ref{Compar.n.fig}.
\\

For procedure {\bf P1}, the increase of $d_{{\rm P1}}^{h}$ is faster and of
order $(h+1)^{4}$. This may be the price to pay when we do not know
the underlying distribution.
\\

One of the main conclusions of this study is that the use of
procedure {\bf P1} needs huge values of $d$ when $h$ is large. For example,
when $h=6$, around 30000 observations are needed to get a power equals
to $95 \%$. This result should be taken into account when one applies
the method to real data sets. 

If one restricts the test to values of $h$ smaller than 6, then
Figure~\ref{Compar.n.fig} shows that the growths of  $d_{{\rm
    P1}}^{h}$ and $d_{{\rm P}} ^{h}$ are of the same order,
$(h+1)^{3.12}$. 
\\

\paragraph{Other non-parametric procedures}
This section highlights the difficulty of testing $k$-monotonicity in
a non-parametric setting when
$k$ increases. Indeed, our conclusions are limited
to the comparison with parametric testing under Poisson
distributions. Morerover, other non parametric procedures could be
used.  For example, we could consider the least-squares estimator of $p$
under the constraint of $k$-monotonicity \cite{giguelay2017estimation} and reject
$H^{k}$ if the distance between this estimator and the empirical distribution
is large, similarly to the tests proposed by \cite{akakpo2014testing} for the discrete monotonicity constraint and \cite{balabdaoui2017testing} for the  discrete convex constraint.
\\

Let us compare our method to the one proposed by \cite{balabdaoui2017testing}, on the basis of their simulation
study. They considered four distributions
\begin{eqnarray*}
p_{0}^{(1)}  & = & Q^{2}_{5} \\
p_{0}^{(2)} &= & \frac{1}{6} Q^{2}_{1} + \frac{1}{6} Q^{2}_{2} +
\frac{1}{3} Q^{2}_{4} + \frac{1}{3} Q^{2}_{5} \\
p_{1}^{(1)}  & = & \cP(\lambda=1.5) \\
p_{1}^{(2)}  & = &  Q^{2}_{5} + 0.008 \delta_{0} - 0.008 \delta_{1},
\end{eqnarray*}
where $\delta_{j}$ is the Dirac distribution in $j$.
For each of these distributions they estimated the rejection
probabilities   on the basis of 500 runs. Their testing  procedure depends on the choice of a tuning parameter
and we report in
Table~\ref{SimFadoua.tb} the results for the best choice of this tuning parameter
(see Table 1 in ~\cite{balabdaoui2017testing}), as well as the results
we get for testing $k=2$ with our Procedure {\bf P1}.
\\

\begin{table}
\begin{center}
\begin{small}
\begin{tabular}{c|cccc|cccc|cccc}
& \multicolumn{4}{c|}{For $d=500$} &\multicolumn{4}{c|}{For $d=5000$}
  &\multicolumn{4}{c}{For $d=50000$}  \\ \hline
 & $p_{0}^{(1)}$ & $p_{0}^{(2)}$ &  $p_{1}^{(1)}$ & $p_{1}^{(2)}$ & 
   $p_{0}^{(1)}$ & $p_{0}^{(2)}$ &  $p_{1}^{(1)}$ & $p_{1}^{(2)}$ &
   $p_{0}^{(1)}$ & $p_{0}^{(2)}$ &  $p_{1}^{(1)}$ & $p_{1}^{(2)}$\\ \hline
B. et al. & 0.054 & 0.020 & 1     & 0.038 & 
            0.062 & 0.018 &  1    & 0.082 & 
            0.05  & 0.016 & 1 & 0.63 \\
 {\bf P1} & 0.046 & 0.034 & 0.956 & 0.034 &
            0.052 & 0.040 & 1 & 0.07&
            0.032 & 0.048 & 1 & 0.36
\end{tabular}
\end{small}
\end{center}
\caption{\label{SimFadoua.tb} Comparison of the procedure proposed by
  Balabdaoui et al.  and Procedure {\bf P1} for testing convexity.}
\end{table}

It appears that Procedure  {\bf P1} is less powerfull than the
procedure based of the asymptotic distribution of the distance between $f$
and its projection  on the space of convex densities. This suggests
that a generalization of such a procedure for testing the
$k$-monotonicity could outperform our procedure based only on the empirical
distribution.\\
Nevertheless this is a rather difficult problem linked to the asymptotic
distribution of the constraint least-squares estimator under shape constraint.
The first difficulty concerns the characterization of the limit
distribution. In fact, on the one hand the limit distributions is not
gaussian -it is  characterized by functions of brownian processes or envelope-type processes- and on the other hand the estimators are not explicit in general (see \cite{Groeneboom2014nonparametric}, Preface). For example, \cite{balabdaoui2010estimation} showed that the limit distribution of the least-squares estimator of a $k$-monotone continuous distribution is a function of the primitives of a two-sided brownian bridge. 
\\
The second difficulty concerns the computation of an approximation of
the limit distribution under the null hypothesis that $p$ is
$k$-monotone. In particular inconsistency of the $k$-knots (the integers
$j$ such that $\nabla^{k}p_{j}$ are strictly positive) may arise in the discrete case. Moreover \cite{balabdaoui2015marshall} pointed out that working with sums instead of Lebesgue measure makes it more difficult to compute the limit distribution. Several authors still managed to compute an approximation of the limit distribution, \cite{rao1969estimation}
in the monotone case, \cite{balabdaoui2017asymptotics} in the convex case
and \cite{balabdaoui2013asymptotics} in the log-concave case for
example. In the convex case, the authors proposed a thresholding
parameter to overcome inconsistency at the knots. It is likely that
the same kind of difficulty should arise concerning  the limit distribution of the least-squares estimator under discrete $k$-monotonicity.

\subsection{\label{SimSplines.st}Simulation for Spline distributions}

As explained in Section~\ref{kmono.st}, any $h$-monotone discrete
distribution $p$ can be decomposed into a mixture of Spline
distributions, see Equations~\eref{QkDec.eq} to~\eref{Qk.eq}. 
\\

\begin{figure}
\begin{center}
\includegraphics[height=.80\textwidth, width=.80\textwidth,angle=0]{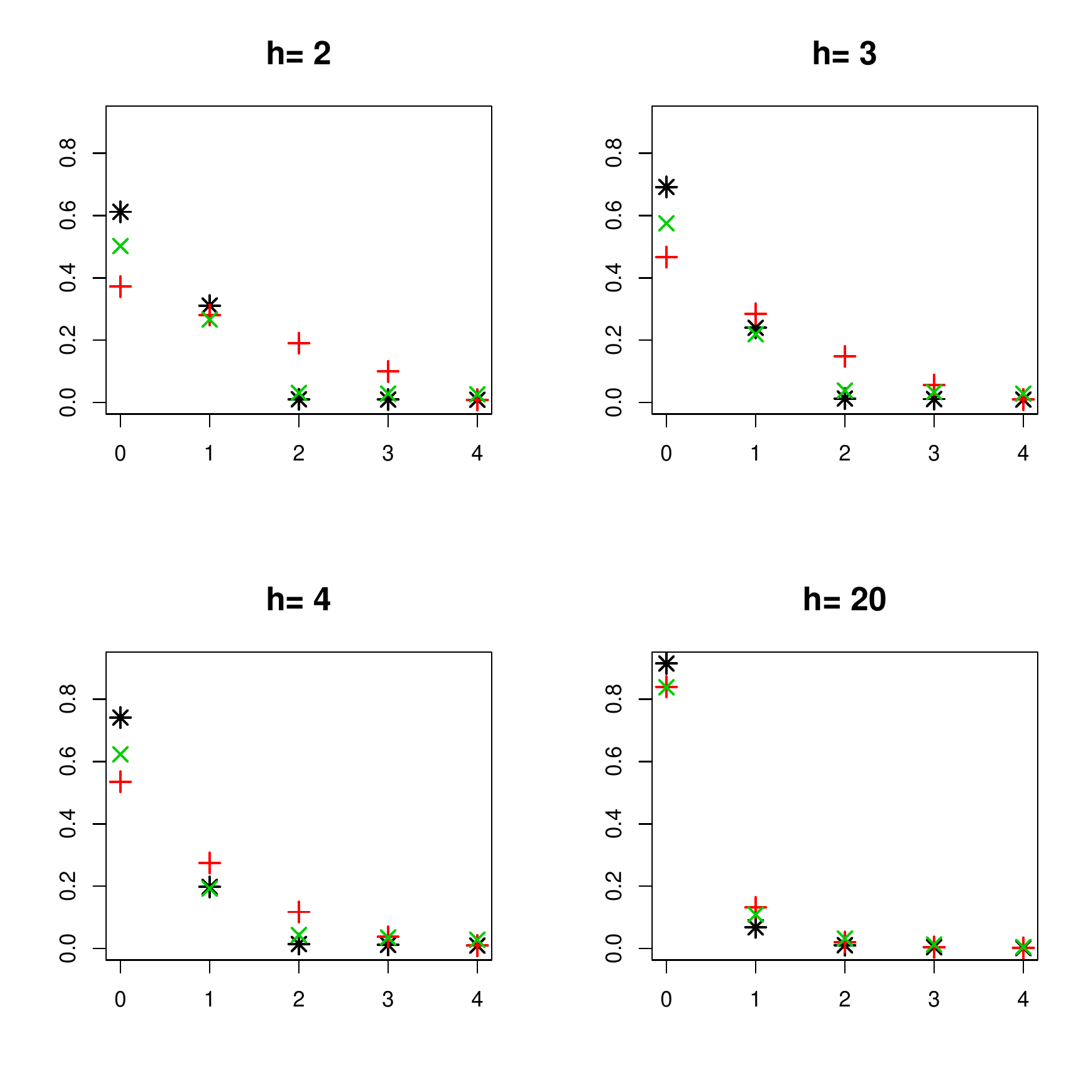}
\end{center}
\caption{\label{SpDistrib.fg} Spline Distributions: * for $0.9
  Q_{1}^{h} + 0.1 Q_{15}^{h}$, {\color{red}$+$} for $0.9  Q_{3}^{h} +
  0.1 Q_{15}^{h}$, {\color{green}x} for $0.7  Q_{1}^{h} +
  0.2 Q_{15}^{h}$.}
\end{figure}

We first consider Spline distributions of
degree $h \in \left\{1, \ldots, 6, 10, 20\right\}$, with one knot in
$\tau$, for $\tau=15$, say $Q^{h}_{15}$. Next we consider
Splines of degree $h$ with two knots, precisely the distributions
\begin{eqnarray*}
 &&  0.9 Q^{h}_{1} + 0.1 Q^{h}_{15} \\
 &&  0.9 Q^{h}_{3} + 0.1 Q^{h}_{15} \\
 &&  0.7 Q^{h}_{1} + 0.3 Q^{h}_{15}
\end{eqnarray*} 
represented at Figure~\ref{SpDistrib.fg}.

\paragraph{Spline distribution with only one knot} 
~
\\
The results for distributions $Q^{h}_{15}$ (not reported) show that it is quite impossible to reject the null
hypothesis $H^{k}$ when considering Spline distribution with one knot
in $\tau = 15$, at least for reasonable values of $d$. Some simple
calculation may help to understand this poor performance. Let us
consider the test of the hypothesis $H^{k}$ for $k=h+1$. Indeed, if
$p=Q^{h}_{\tau}$, then $\nabla^{h+1} p_{j} < 0 $ for $j=\tau-1$ only,
and 
\begin{equation*}
 \nabla^{h+1} p_{\tau-1} = p_{\tau-1}- (h+1)
p_{\tau} = -p_{\tau}.
\end{equation*}
The standard-error of its empirical estimator,
$\nabla^{h+1} f_{\tau-1}$,
may be approximated by 
\begin{equation*}
 \sqrt{p_{\tau-1} + (h+1)^2 p_{\tau}}/\sqrt{d}.
\end{equation*}
Let us consider the  test of
the single  hypothesis ``$\nabla^{h+1} p_{\tau-1} \geq  0 $'': using the
Gaussian approximation, the null hypothesis will be rejected if 
\begin{equation*}
 \sqrt{d} \nabla^{h+1} f_{\tau-1} < \nu_{\alpha} \sqrt{
p_{\tau-1}- (h+1)p_{\tau}}.
\end{equation*}
Replacing $\nabla^{h+1} f_{\tau-1}$ by $\nabla^{h+1} p_{\tau-1}$, it
appears that $d$ should satisfy 
\begin{equation}
 d > \frac{\nu^{2}_{\alpha}(p_{\tau-1} + (h+1)^2 p_{\tau})
 }{p^{2}_{\tau}} = \nu^{2}_{\alpha} C_{h+\tau}^{h} \left(h +
 (h+1)^{2}\right),
\label{dreject.eq}
\end{equation}
in order to reject ``$\nabla^{h+1} p_{\tau-1} \geq  0 $''. Clearly $d$
increases with $h$,   and $\tau$ (see Table~\ref{dreject.tb}).
\\

\begin{table}
\begin{center}
\begin{tabular}{cccccc}
 $h = 1$ & $h = 2$ &  $h = 3$ &  $h = 4$ &   $h = 5$ &   $h = 6$ \\
   200 & 4000 & 42000 & 304000 & 1720000 & 8080000
\end{tabular}
\end{center}
\caption{\label{dreject.tb}
Minimum value of $d$ satisfying Equation~\eref{dreject.eq} for
$\tau=15$ and $\alpha=0.05$.}
\end{table}

In practical situations, the distribution $p$ is unknown, and the test
of $H^{h+1}$ lies on 
a multiple testing procedure, making even more difficult to reject
$H^{h+1}$. 

\paragraph{Spline distributions with 2 knots}
~
\\
The results for distributions of the form $\pi
Q^{h}_{\ell} + (1-\pi) Q^{h}_{\tau}$ are given in
Tables~\ref{SimSplineMod1.tab} to~\ref{SimSplineMod3.tab}. We  report the estimated probabilities of rejection for the
test of the hypothesis $H^{k}$ for $k=h$ in order to estimate the
level of the test, and $k=h+1$ to estimate the power. 
\\

The level of the tests are nearly equal to $5\%$. The power decreases
with $h$ for all models and procedures and is greater for  a
mixture of spline distributions 
such that the first knot is close to  0, and such that the mass in the
first knot is large. 
Nevertheless, procedure {\bf
  P1} gives the best results for the first and third models where the
first knot appears in $j=1$, while procedure {\bf P2} performs better
for the second model. 
\\

When $h$ equals 1 or 2, the power of the test is close to one for the
first and third models for $d=1000$. For the second model, $d=5000$ is
needed to get such a power. When $h$ increases, for example $h=5$, the
difficulty for testing $H^{h+1}$ for the second model is confirmed:
for $d=30000$, the power remains smaller than $10\%$.

\begin{table}[htb]
\begin{center}

$d=1000$
\medskip

\begin{tabular}{c|l|cccccc}
         &       & $h=6$ & $h=5$ & $h=4$ & $h=3$ & $h=2$ &$h=1$ \\ \hline
{\bf P1} & $k=h$ & 0.042 & 0.050 & 0.044 & 0.062 & 0.052 & 0.064 \\
         &$k=h+1$& 0.124 & 0.278 & 0.630 & 0.936 & 1.000 & 1 \\\hline
{\bf P2} & $k=h$ & 0.014 & 0.036 & 0.032 & 0.018 & 0.014 & 0.040 \\
         &$k=h+1$& 0.034 & 0.106 & 0.252 & 0.712 & 0.996 & 1.000  \\\hline
\end{tabular}
\medskip

$d=5000$
\medskip

\begin{tabular}{c|l|cccccc}
         &       & $h=6$ & $h=5$ & $h=4$ & $h=3$ & $h=2$ &$h=1$ \\\hline
{\bf P1} & $k=h$ & 0.034 & 0.040 & 0.060 & 0.052 & 0.050 & 0.060 \\
         &$k=h+1$& 0.360 & 0.816 & 0.990 & 1 & 1 & 1\\\hline
{\bf P2} & $k=h$ & 0.028 & 0.046 & 0.032 & 0.032 & 0.020 & 0.038 \\
         &$k=h+1$& 0.126 & 0.566 & 0.952 & 1     & 1     & 1\\\hline
\end{tabular}
\medskip

$d=30000$
\medskip

\begin{tabular}{c|l|cccccc}
         &       & $h=6$ & $h=5$ & $h=4$ & $h=3$ & $h=2$ &$h=1$ \\\hline
{\bf P1} & $k=h$ & 0.040 & 0.048 & 0.046 & 0.040 & 0.044 & 0.052 \\
         &$k=h+1$& 0.964 & 1     & 1     & 1     & 1     & 1\\\hline
{\bf P2} & $k=h$ & 0.032 & 0.030 & 0.018 & 0.030 & 0.056 & 0.038 \\
         &$k=h+1$& 0.818 & 1     & 1     & 1     & 1     & 1\\\hline
\end{tabular}
\end{center}
\caption{\label{SimSplineMod1.tab} 
  Estimated probabilities of  rejecting  the
hypothesis $H^{k}$, for Spline distribution $p^{h}=0.9 Q^{h}_{1} + 0.1
 Q^{h}_{15}$.}
\end{table}

\begin{table}[htb]
\begin{center}

$d=1000$
\medskip

\begin{tabular}{c|l|cccccc}
         &       & $h=6$ & $h=5$ & $h=4$ & $h=3$ & $h=2$ &$h=1$ \\\hline
{\bf P1} & $k=h$ & 0.052 & 0.060 & 0.038 & 0.058 & 0.046 & 0.038 \\
         &$k=h+1$& 0.044 & 0.056 & 0.050 & 0.048 & 0.212 & 1\\\hline
{\bf P2} & $k=h$ & 0.034 & 0.018 & 0.010 & 0.026 & 0.004 & 0.030 \\
         &$k=h+1$& 0.030 & 0.022 & 0.058 & 0.100 & 0.508 & 1\\\hline
\end{tabular}

\medskip

$d=5000$
\medskip

\begin{tabular}{c|l|cccccc}
         &       & $h=6$ & $h=5$ & $h=4$ & $h=3$ & $h=2$ &$h=1$ \\\hline
{\bf P1} & $k=h$ & 0.056 & 0.058 & 0.040 & 0.054 & 0.058 & 0.054\\
         &$k=h+1$& 0.060 & 0.048 & 0.038 & 0.070 & 0.998 & 1\\\hline
{\bf P2} & $k=h$ & 0.028 & 0.034 & 0.036 & 0.018 & 0.022 & 0.050 \\
         &$k=h+1$& 0.042 & 0.050 & 0.080 & 0.496 & 1     & 1\\\hline
\end{tabular}
\medskip

$d=30000$
\medskip

\begin{tabular}{c|l|cccccc}
         &       & $h=6$ & $h=5$ & $h=4$ & $h=3$ & $h=2$ &$h=1$ \\\hline
{\bf P1} & $k=h$ & 0.040 & 0.046 & 0.052 & 0.044 & 0.034 & 0.046\\
         &$k=h+1$& 0.040 & 0.046 & 0.060 & 0.964 & 1     & 1\\\hline
{\bf P2} & $k=h$ & 0.036 & 0.044 & 0.044 & 0.034 & 0.046 & 0.058 \\
         &$k=h+1$& 0.042 & 0.088 & 0.518 & 1     & 1     & 1\\\hline
\end{tabular}
\end{center}
\caption{\label{SimSplineMod2.tab} 
  Estimated probabilities of  rejecting  the
hypothesis $H^{k}$, for Spline distribution $p^{h}=0.9 Q^{h}_{3} + 0.1
 Q^{h}_{15}$.}
\end{table}

\begin{table}[htb]
\begin{center}

$d=1000$
\medskip

\begin{tabular}{c|l|cccccc}
         &       & $h=6$ & $h=5$ & $h=4$ & $h=3$ & $h=2$ &$h=1$ \\\hline
{\bf P1} & $k=h$ & 0.048 & 0.040 & 0.048 & 0.052 & 0.068 & 0.050 \\
         &$k=h+1$& 0.072 & 0.106 & 0.298 & 0.716 & 0.998 & 1\\\hline
{\bf P2} & $k=h$ & 0.028 & 0.022 & 0.046 & 0.024 & 0.030 & 0.052 \\
         &$k=h+1$& 0.028 & 0.034 & 0.126 & 0.354 & 0.938 & 1\\\hline
\end{tabular}

\medskip

$d=5000$
\medskip

\begin{tabular}{c|l|cccccc}
         &       & $h=6$ & $h=5$ & $h=4$ & $h=3$ & $h=2$ &$h=1$ \\\hline
{\bf P1} & $k=h$ & 0.036 & 0.040 & 0.048 & 0.052 & 0.054 & 0.030\\
         &$k=h+1$& 0.132 & 0.312 & 0.802 &     1 &     1 &     1\\\hline
{\bf P2} & $k=h$ & 0.022 & 0.042 & 0.036 & 0.034 & 0.036 & 0.026 \\
         &$k=h+1$& 0.042 & 0.134 & 0.492 & 0.990 &     1 & 1\\\hline
\end{tabular}
\medskip

$d=30000$
\medskip

\begin{tabular}{c|l|cccccc}
         &       & $h=6$ & $h=5$ & $h=4$ & $h=3$ & $h=2$ &$h=1$ \\\hline
{\bf P1} & $k=h$ & 0.036 & 0.026 & 0.052 & 0.032 & 0.046 & 0.054 \\
         &$k=h+1$& 0.414 & 0.942 &     1 &     1 &     1 & 1\\\hline
{\bf P2} & $k=h$ & 0.042 & 0.030 & 0.024 & 0.044 & 0.038 & 0.058 \\
         &$k=h+1$& 0.180 & 0.770 &     1 &     1 &     1 & 1 \\\hline
\end{tabular}
\end{center}
\caption{\label{SimSplineMod3.tab} 
  Estimated probabilities of  rejecting  the
hypothesis $H^{k}$, for Spline distribution $p^{h}=0.7 Q^{h}_{1} + 0.3
 Q^{h}_{15}$.}
\end{table}

\subsection{\label{boot.st}Comparison with the bootstrap procedure}

Let us describe the bootstrap procedure for estimating  the quantities
$q_{\alpha}^{k}$ and $u_{\alpha}^{k}$. Let  $(X_{1}^{*}, \ldots, X_{d}^{*})$ be a $d$-sample
distributed according to the empirical distribution of $(X_{1}, \ldots, X_{d})$,
and let $f_{j}^{*}, j=0, \ldots, \tau^{*}$ be the empirical  estimator
of the bootstrap distribution. Then
\begin{equation*}
 q^{* k}_{\alpha} = \inf_{q} \left\{\P_{X} \left( 
\min_{0  \leq j \leq \widehat{\tau}-1}
\sqrt{d} \nabla^{k} (f_{j}^{*} - f_{j}) \leq q
\right)\right\},
\end{equation*}
where $\P_{X}$ denotes the  conditional distribution given $(X_{1},
\ldots, X_{d})$.

For estimating $u^{k}_{\alpha}$ we use a double bootstrap. For a given
$u$ and for $0\leq j \leq \widehat{\tau}-1$, let $\nu^{*}_{j,u}$ be
defined as follows:
\begin{equation*}
 \nu^{*}_{j,u} = \inf_{\nu} \left\{
\P_{X} \left( \sqrt{d} \nabla^{k} (f_{j}^{*} - f_{j}) \leq \nu 
\sqrt{A_{j}^{k T} \widehat{\Gamma} A_{j}^{k}} \right) = u
\right\}
\end{equation*}
Next let $(X_{1}^{**}, \ldots, X_{d}^{**})$ be a $d$-sample
distributed according to the empirical distribution of $(X_{1},
\ldots, X_{d})$, independant of $(X_{1}^{*}, \ldots, X_{d}^{*})$, and
let $f_{j}^{**}, j=1, \ldots, \tau^{**}$ be the empirical
frequencies. The bootstrap estimator of $u_{\alpha}$ is defined as
follows:
\begin{equation*}
u^{* k}_{\alpha}  = \max_{0 \leq u \leq 1} \left\{
\P_{X} \left( 
\min_{0 \leq j \leq \widehat{\tau}-1} \left\{
\sqrt{d} \nabla^{k} (f_{j}^{**} - f_{j}) - \nu^{*}_{j,u} \sqrt{A_{j}^{k T} \widehat{\Gamma} A_{j}^{k}}
\right\} \leq 0
\right) = \alpha
\right\}.
\end{equation*}

The results (not shown) are equivalent to those of procedures {\bf P1} and {\bf
  P2}. 

Although the validity of the boostrap procedure, like our
  procedures {\bf P1} and {\bf P2}, lies on asymptotic arguments,  we could have
  expected a different  behaviour 
  of the bootstrap procedure, because bootstrap does not use the
  approximation of 
  the empirical distribution by the Gaussian distribution for
  practical calculation. This is
  clearly not the case, may be because our simulation study consider 
  values of the sample size $d$ large enough to guarantee  that  the distribution of the empirical
  frequencies is closed to the Gaussian approximation.

\section{\label{Estk.st}Estimating the degree of monotonicity of $p$}

\subsection{Estimator and asymptotic properties}

We propose a procedure for estimating $k$, the degree of
monotonicity of $p$, based on the testing procedures described in
the previous section.

For some  $\alpha$ and  $k_{\max}$, we define $\widehat{k}_{\alpha}$ as  follows

\begin{itemize}
\item if there exists $1 \leq \ell \leq k_{\max}$ such that $H^{\ell}$
  is rejected, then
\begin{equation*} 
\widehat{k}_{\alpha} = \inf_{1 \leq \ell \leq k_{\max}} \left\{ H^{\ell} \mbox{ is rejected at
level }\alpha\right\} -1
\end{equation*}
\item if not, $\widehat{k}_{\alpha} = k_{\max}$.
\end{itemize}

We show that $\widehat{k}_{\alpha}$ is asymptotically close to $k$.

\begin{theo}
\label{kChap.th}
Let $p$ be a $k$-monotone distribution and let  $\widehat{k}_{\alpha}$
be defined as above. For all $\ell\geq 1$, let $\sigma^{\ell}$ and
$\zeta^{\ell}$ be defined as in Theorem~\ref{TestkLevel.th}. 
According to the testing procedure for
calculating $\widehat{k}$, let us assume that  the following property
is statisfied: 
\begin{description}
\item[P1] If for all $1\leq \ell \leq k-2$,
\begin{equation*}
 \min_{0\leq j \leq
    \tau-1} \nabla^{\ell} p_{j} \geq \sqrt{\frac{2}{d}} \sigma^{\ell}
\sqrt{\log(\tau) + \frac{1}{2} \log(d)},
\end{equation*}
\item[P2] If for all $1\leq \ell \leq k-2$, 
\begin{equation}
 \min_{0\leq j \leq
    \tau-1} \nabla^{\ell} p_{j} \geq \sqrt{\frac{2}{d}} \zeta^{\ell}
\sqrt{\log(\tau) + \frac{1}{2} \log(d)}, \label{pbssestim.eq}
\end{equation}
then
\begin{equation*}
\lim_{d \rightarrow \infty} \P\left( \widehat{k}_{\alpha} \leq k-1
\right) \leq \alpha 
\end{equation*}
\end{description}
If $k\leq k_{\max}-1$ and if $p$  satisfies the following
property: 
\begin{description}
\item[P1] 
\begin{equation}
 \exists j_{0},    \nabla^{k+1}p_{j_{0}} + \frac{1}{\sqrt{d}}  
\left(\sigma^{k} \sqrt{\log\frac{\tau}{\alpha}}
 +\zeta^{k}_{j_{0}} \sqrt{-2 \log
   \frac{k_{\max}-k}{\sqrt{d}}}\right) \leq 0,
\label{far.eq}
\end{equation}
\item[P2] 
\begin{equation}
 \exists j_{0},    \nabla^{k+1}p_{j_{0}} + \frac{1}{\sqrt{d}}  
\left( \sqrt{\log\frac{\tau}{\alpha}}
 + \sqrt{-2 \log
   \frac{k_{\max}-k}{\sqrt{d}}}\right)\zeta^{k+1}_{j_{0}} \leq 0,
\end{equation}
\end{description}
then 
\begin{equation*}
\lim_{d \rightarrow \infty} \P\left( \widehat{k}_{\alpha} \geq k+1
\right) = 0.
\label{far.eq}
\end{equation*}
\end{theo}

This theorem, shown in Section~\ref{proofs-tests.st}, claims that if
$p$ is $k$-monotone, then the probability that $\widehat{k}_{\alpha}
\leq k-1$ is asymptotically smaller than $\alpha$. Moreover if $p$ is
far enough from $(k+1)$-monotone densities, then the probability that
$\widehat{k}_{\alpha} \geq k+1$ tends to zero. 

\subsection{Simulation study}

The properties of $\widehat{k}$ for $k_{\max}=6$ are assessed on the
basis of the simulation study presented before. The results are given  at
Tables~\ref{SimulkChapPoissonP1.tab}
  to~\ref{SimulkChapSplineMod2.tab}. They are reported for models
  whose degree of monotonicity is smaller than 5, when using the procedure
  that proved to maximise the power in the simulation study presented
  in the previous Section. 

Let $h$ be the true degree of
  monotonicity of the distribution $p$. From these results, we deduce that
\begin{itemize}
\item Probability to underestimate $h$ when $p$ is $h$-monotone.

The estimator $\widehat{k}_{\alpha}$ equals $h-1$ in nearly $5\%$ of the
  runs. When $h \geq 2$,  the
  number of runs for which $\widehat{k}_{\alpha} \leq h-2$ is 0. This
  result confirms the first part of Theorem~\ref{kChap.th}, see
  Equation~\eref{pbssestim.eq}.

\item Probability to over estimate $h$ when $p$ is $h$-monotone. 

This probability is linked with the power of the test: if the test has
a low power, the degree of monotonicity will be overestimated. When $h$
increases, the probability to get $\widehat{k}_{\alpha} \geq h+1$, and
in particular $\widehat{k}_{\alpha} = k_{\max}$,
increases. This overestimation decreases with $d$. For the spline
distributions, if $d=5000$ the results are correct for $h \leq 3$. 
\end{itemize}

\begin{table}[htb]
\begin{center}
\begin{small}
\begin{tabular}{r|rrrrr||rrrrr||rrrrr}
\multicolumn{1}{c}{}& \multicolumn{5}{c||}{$d=1000$} & \multicolumn{5}{c||}{$d=5000$}
& \multicolumn{5}{c}{$d=30000$} \\
$h$& $5$ & $4$ & $3$ & $2$ & $1$ &$5$ & $4$ & $3$ & $2$ & $1$ &$5$ &
$4$ & $3$ & $2$ & $1$ \\ \hline
$\widehat{k}$ & 5.77 & 5.27 & 4.42 & 2.68 & 0.99 &5.50 &4.30& 3.00&
1.94 &0.92&4.97 &3.96 &2.93 &1.96 &0.97\\\hline
0 & 0 & 0 & 0 & 0 & 2  0&0&0&0&0&2.8 & 0&0&0&0&3.2\\
1 & 0 & 0 & 0 & 3.4 & 97.6 & 0&0&0& 5.6& 97.2 & 0&0&0& 4.4&96.8\\
2 & 0 & 0 & 4 & 74.8 & 0.2& 0&0&5.6&94.4&0 & 0&0& 7.2&95.6&0\\
3 & 0& 4.4& 36.6& 5.4 &0&0&4.2&90&0&0  & 0&4.4&92.8&0&0\\
4 & 5 & 24.8 &16& 0&0&4&70.8&3.6&0&0& 4.4&95.6&0&0&0\\
5 & 12.6& 10&  0.6 &0 & 0 & 41.8 & 16.2 & 0&0&0&94.6&0&0&0&0 \\
6 & 82.4& 60.8& 42.8 &16.4 &0.2& 54.2&8.8&0.8&0&0&1&0&0&0&0\\
\end{tabular}
\end{small}
\end{center}
\caption{\label{SimulkChapPoissonP1.tab} 
For Poisson 
  distributions with parameters $\lambda^{h}$ given at
  Table~\ref{lambdaValues.tab}: Estimated degree of monotonicity with
  procedure {\bf P1} and $\alpha=5\%$. For each value of $d$, the first row
  of the table
  gives $h$, the second the mean of
  $\widehat{k}_{\alpha}$ estimated over 500 runs, the following rows
  give the histogram of the 
  estimated values of $\widehat{k}_{\alpha}$ (as $100 \times$ percentages).}
\end{table}

\begin{table}[htb]
\begin{center}
\begin{small}
\begin{tabular}{r|rrrrr||rrrrr||rrrrr}
\multicolumn{1}{c}{}& \multicolumn{5}{c||}{$d=1000$} & \multicolumn{5}{c||}{$d=5000$}
& \multicolumn{5}{c}{$d=30000$} \\
$h$& $5$ & $4$ & $3$ & $2$ & $1$ &$5$ & $4$ & $3$ & $2$ & $1$ &$5$ &
$4$ & $3$ & $2$ & $1$ \\ \hline
$\widehat{k}$ & 5.67 & 4.48 & 3.00 & 1.95 & 0.94 &
5.14 & 3.95 & 2.95 & 1.95 &0.94& 4.95 & 3.95 & 2.96& 1.96 & 0.95
\\\hline
0 & 0 & 0 & 0 & 0 & 6.4 & 0 & 0 & 0 & 0 & 6 & 0 & 0 & 0 & 0 & 5.2  \\
1 & 0 & 0 & 0 & 5.2 & 93.6 & 0 & 0 & 0 & 5 & 94 & 0 & 0 & 0 & 4.4 & 94.8 \\
2 & 0 & 0 & 6.2 & 94.8 & 0 & 0 & 0 & 5.2 & 95 & 0 & 0 & 0 & 4 & 95.6 & 0 \\
3 & 0 & 4.4 & 87.4 & 0 & 0 & 1.6 & 1.6 & 68.8 & 0.4 & 0 & 0 & 4.6 & 96
& 0 & 0 \\
4 & 5 & 58.6 & 6.4 & 0 & 0 & 1 & 23.2 & 26.8 & 0 & 0 & 4.8 & 95.4 &0 &
0 & 0 \\
5 & 22.8 & 21.2 & 0 & 0 & 0 & 7.2 & 28.2 & 1.6 & 0 & 0 & 95.2 & 0 & 0
& 0 & 0 \\
6 & 72.2 & 15.8 & 0 & 0 & 0 & 87.8 & 45 & 0.4 & 0 & 0 &0& 0 & 0
& 0 & 0 \\
\end{tabular}
\end{small}
\end{center}
\caption{\label{SimulkChapSplineMod1.tab} 
For Spline  
  distributions with two knots defined as $0.9Q^{h}_{1}+ 0.1 Q^{h}_{15}$:
  Estimated degree of monotonicity with procedure {\bf P1} and $\alpha=5\%$. For each value of $d$, the first row
  of the table
  gives $h$, the second the mean of
  $\widehat{k}_{\alpha}$ estimated over 500 runs, the following rows
  give the histogram of the 
  estimated values of $\widehat{k}_{\alpha}$ (as $100 \times$ percentages).}
\end{table}

\begin{table}[htb]
\begin{center}
\begin{small}
\begin{tabular}{r|rrrrr||rrrrr||rrrrr}
\multicolumn{1}{c}{}& \multicolumn{5}{c||}{$d=1000$} & \multicolumn{5}{c||}{$d=5000$}
& \multicolumn{5}{c}{$d=30000$} \\
$h$& $5$ & $4$ & $3$ & $2$ & $1$ &$5$ & $4$ & $3$ & $2$ & $1$ &$5$ &
$4$ & $3$ & $2$ & $1$ \\ \hline
$\widehat{k}$ & 5.90 & 5.78 & 5.47 & 2.71 & 0.97 & 5.83 & 5.73 & 3.89
& 1.97 & 0.95 & 5.11 & 3.95 & 2.94 & 1.96 & 0.94 
\\\hline
0 & 0.2 & 0.8 & 1.2 & 0.4 & 3 & 0   & 0   & 0   & 0  &4.6  &  0 & 0   & 0   & 0   & 5.8  
\\
1 & 0   & 0   & 0   & 0.4 & 97& 0   & 0.6   & 1.6   & 3.2 & 95.4&  0 & 0.4 & 1.6 & 3.8 &94.2 
\\
2 & 0.2 & 0.8 & 2.4 & 50.4& 0 & 2.2 & 1.8 & 1.4 & 96.8& 0  & 2.2& 1.2 & 3   & 96.2& 0 
\\
3 & 2.4 & 3   & 7   & 39.2& 0 & 1.6 & 1.8 & 97  & 0   & 0  & 1.8&  1  & 95.4& 0   & 0 
\\
4 & 0.2 & 2.4 & 6.8 &2.8  & 0 & 0.6 & 80.6& 0   & 0   & 0  & 0.6& 97.4&0    &0    & 0
\\
5 & 0.8 &0.4  & 2   & 0.2 & 0 & 18  & 11.8& 0   & 0   & 0  &73.4& 0   & 0   & 0   & 0
\\
6 & 96.2& 92.6& 81.1& 6.6 & 0 & 77.6&  3.4& 0   & 0   & 0  &22 & 0   & 0   & 0   & 0 \\
\end{tabular}
\end{small}
\end{center}
\caption{\label{SimulkChapSplineMod2.tab} 
For Spline  
  distributions with two knots defined as $0.9Q^{h}_{3}+ 0.1 Q^{h}_{15}$:
  Estimated degree of monotonicity with procedure {\bf P2} and $\alpha=5\%$. For each value of $d$, the first row
  of the table
  gives $h$, the second the mean of
  $\widehat{k}_{\alpha}$ estimated over 500 runs, the following rows
  give the histogram of the 
  estimated values of $\widehat{k}_{\alpha}$ (as $100 \times$
  percentages).}
\end{table}

\section{Number of classes in a population}
\label{Model.st}

Let us now consider the case where the total number of classes in a
population is unknown, and where we aim at estimating this number
based on the abundances that are observed for a series of classes. The
problem is then to estimate the number of unobserved classes. 

This problem was first raised in the context of ecology for estimating
species richness of a population and  traces back to Fisher et al.~\cite{fisher1943relation}. Nevertheless it also occurs in a
wide variety of domains, as in social and medical
sciences, epidemiology, computer
science, \ldots.  Since the  contribution of
Fisher et al., many
publications have considered this problem proposing different
statistical modelings and estimators. A presentation of these
different approaches was given by Bunge and Fitzpatrick~\cite{bunge1993estimating} for
example. A more recent short review can be found
in~\cite{durot2015nonparametric}, see also~\cite{böhning2017capture}.

In this section, we first
describe the observations and the statistical modeling,  making thus
the link between the $k$-monotonicity of the abundance distribution of
the classes, and the estimator of the number of total classes. Then we
carry out a simulation study in order to assess the properties of our
estimator, and finally we consider three real case studies.

\subsection{The observations}\label{statprob.st}

Suppose that the
population is composed of $N$ classes and for $i = 1 \dots N$, denote by
$A_{i}$ the abundance (that is the number of observed individuals) of
class $i$ and by $S_{j}$ the number of classes with abundance
$j$ in a sample. The total number of observed classes is $D = \sum_{j \geq 1}
S_j$ whereas $S_0$ is the number of unobserved classes. The
total number of classes is $N = S_0 + D$ and, because $D$ is observed,
the estimation of $N$ amounts to the estimation of $S_0$. We
will denote by $n$ the sample size: $n = 
\sum_i A_i = \sum_j j S_j$.

We assume that the $A_{i}$'s are independent
variables with the same distribution $p=(p_{0},p_{1},\dots,p_{n})$,
called the {\sl abundance distribution}.

As only classes that are present in the sample can be
counted,  classes for which $A_{i}=0$ are not observed. Thus, we only
observe the zero-truncated counts $X_{1},\dots,X_{D}$, where $X_{i}$
is the abundance of the $i$-th observed classes in the sample. As it
is shown by~\cite{durot2015nonparametric} (lemma 1 of the on line supporting information), $D\sim\mbox{Bin}(N,1-p_{0})$, and conditionally on $D$,
$X_{1},\dots,X_{D}$ are i.i.d. random variables with distribution
$p^+$ defined by 
\begin{equation}\label{p+.eq}
p^{+}_j= \frac{p_j}{1-p_0},
\mbox{ for all integers } j\geq 1.
\end{equation}

Therefore we propose to estimate $N$ by 
\begin{equation}
\widehat{N} = \frac{D}{1-\widehat{p}_{0}},
\label{Nhat.eq}
\end{equation}
where $\widehat{p}_{0}$ is an estimator of $p_{0}$. 

The problem comes to estimate $p_{0}$. As we observe $X_{1},\dots,X_{D}$ from distribution $p^{+}$, we are
able to estimate $p^{+}$. Nevertheless, identifiability
conditions are needed to infer $p_{0}$ from the estimation of
$p^{+}$. This is the object of the following section.

\subsection{\label{kmono.st}The assumption of a $k$-monotone abundance distribution}\label{CvxAb.st}
To make $p_{0}$, and thus 
$N$, identifiable, we propose a
nonparametric modeling of $p$, assuming that $p$  is a discrete $k$-monotone
abundance distribution, as defined in Section~\ref{Testk.st}. In
particular, we know that $p$ is written as a mixture of distribution
$Q_{\ell}^{k}$: for all $j \in \N$, $p_{j} = \sum_{\ell\geq
  0} \pi_{\ell}^{k} Q_{\ell}^{k}(j)$.

Our interpretation of this mixture is that the
set of classes is separated into groups, each class having probability $\pi^{k}_{\ell}$ to
belong to the group $\ell$ of classes, and the abundance distribution of all classes in the group $\ell$ is the 
distribution $Q^{k}_\ell$. As the first component $Q_0^{k}$ is a Dirac mass at 0, it refers to
classes for which the only abundance that could be observed is 0. This group
simply defines absent classes, and therefore $\pi^{k}_{0}$ has to be
zero in an abundance distribution. This leads to the following
definition.

\paragraph{Definition of a $k$-monotone abundance distribution:} {\it
The distribution $p$ on $\N$ is a $k$-monotone abundance distribution if
there exist positive weights $\pi^{k}_{\ell}$ 
satisfying $\sum_{\ell \geq 1} \pi^{k}_{\ell} = 1$, such that $p_j =
\sum_{\ell\geq  1} \pi^{k}_{\ell}Q^{k}_\ell(j)$ for all integers $j\geq 0$.}

In the following, we assume that the abundance distribution $p$ is a
$k$-monotone  abundance distribution. It then follows
from~\eref{pi.eq} that $\pi^{k}_{0} = \nabla^{k}p_0=0$, or equivalently, that
\begin{equation}
\frac{1}{1-p_0}  = 1 - \sum_{h=1}^{k} (-1)^{h} C_{k}^{h}
p^{+}_{h},
\label{theta.eq}
\end{equation}
where $p^{+}$ is the zero-truncated distribution defined by~\eref{p+.eq}.

The distribution $p^+$ is identifiable since we observe
$X_{1},\dots,X_{D}$ which are i.i.d. with distribution $p^+$
conditional on $D$. Therefore,  it follows from \eqref{theta.eq} that
$1-p_{0}$ is identifiable and because $D\sim\mbox{Bin}(N,1-p_{0})$, we
conclude that $N$ also is identifiable. This shows that our assumption
is sufficient to avoid identifiability problems. We will see how to
estimate $p_{0}$ in the following section.

Let us remark that $\nabla^{k}p_0=0$ is equivalent to 
\begin{eqnarray*}
p_{0}& = & - \sum_{h=1}^{k}(-1)^{h} C_{k}^{h} p_{h} \\
& = & - \sum_{h=1}^{k-1}(-1)^{h} C_{k-1}^{h} p_{h} - \nabla^{k}p_{0} +
\nabla^{k-1}p_{0} \\
& = & -\sum_{h=1}^{k-1}(-1)^{h} C_{k-1}^{h} p_{h} + \nabla^{k-1}p_{1},
\end{eqnarray*}
the last equality being deduced from the definition of $\Delta^{k}$
given at Equation~\eref{defDelta.eq}. Therefore if we denote by
$p_{0}^{k}$ the value of $p_{0}$ under the assumption that $p$ is a
$k$-monotone abundance distribution, then
\begin{equation*}
p_{0}^{k}  = p_{0}^{k-1} + \nabla^{k-1}p_{1} > p_{0}^{k-1},
\end{equation*}
because $p$ is strictly $(k-1)$-monotone. Therefore, the mass in 0 of $p$
increases with $k$ when $p$ is assumed to be a $k$-monotone abundance distribution.

\section{Estimating the number of classes}
\label{Estim.st}

In order to estimate $N$, we  first build an estimator for $1/(1-p_0)$
based on Equation~\eref{theta.eq} and then apply
Equation~\eref{Nhat.eq}. 

\subsection{\label{Nchap.st}Estimator based on the relative frequencies} 
For all $j \geq 1$, the empirical estimator (which is the more commonly used estimator for a discrete distribution) of
$p^{+}_{j}$ is $f_j = S_{j}/D$.
Using this estimator in \eqref{Nhat.eq} leads to the estimator
\begin{equation}\label{NEmp.eq}
\widehat{N^{k}} = D  - \sum_{h=1}^{k} (-1)^{h} C_{k}^{h}S_{h}
\end{equation}
Let $\widehat{s}^{k}$ be  defined
as follows
\begin{equation*}
 \widehat{s}^{k} = \sqrt{\sum_{h=1}^{k} \left((-1)^{h+1}+  C_{k}^{h}\right)
C_{k}^{h}
  S_{h}}.
\end{equation*}
If $\widehat{s}^{k} \neq 0$, one can derive from the central limit theorem that
\begin{equation*}
\frac{\widehat{N}_{k}-N}{\widehat{s}^{k}} \mbox{ converges in law to }
\N(0,1).
\end{equation*}

Let us give the following remarks:
\begin{rmk} \label{NsupD.rmk}
If the empirical estimator of $p^{+}$ is far from being
  $k$-monotone, then the quantity $ \sum_{h=1}^{k} (-1)^{h}
  C_{k}^{h}S_{h}$ may be positive.  Clearly the estimator of $N$ is
  expected to be greater than $D$ (or equal). Therefore, for a given $k$, the
  method can be applied only if $ \sum_{h=1}^{k} (-1)^{h}
  C_{k}^{h}S_{h} \leq 0$. This condition will guarantee that
  $\widehat{s}^{k}$ is well defined. For example, if we choose $k=2$, the
  empirical distribution should 
  statisfy $2S_{1}-S_{2}\geq 0$ and $\widehat{s}^{k}= \sqrt{6 S_{1}}$.
\end{rmk}
\begin{rmk}
The bias and variance of $\widehat{N}^{k}$ can be easily
  calculated: (see Section~\ref{biasVar.st})
\begin{eqnarray*}
  \E\left(\widehat{N}^{k} \right) &= &N - N \nabla^{k} p_{0} \\
\V\left( \widehat{N}^{k}/\sqrt{N}\right) &= &p_{0}+
\sum_{h=1}^{k}\left(C_{k}^{h}\right)^{2}p_{h} -
\left(\nabla^{k}p_{0}\right)^{2}
\end{eqnarray*}
If $p$ is a $k$-monotone abundance distribution, then
$\nabla^{k} p_{0} =0$, $\widehat{N}^{k}$ has no bias and 
\begin{equation*}
  \V\left( \widehat{N}^{k}/\sqrt{N}\right) = p_{0}+
\sum_{h=1}^{k}\left(C_{k}^{h}\right)^{2}p_{h} = \sum_{h=0}^{k}\left(C_{k}^{h}\right)^{2}p_{h}. 
\end{equation*}
In that case, the variance of $\widehat{N}^{k}$ increases with $k$. 
\end{rmk}
\begin{rmk}
Let us assume now that  $p$ is a $k$-monotone abundance
  distribution, but we estimate $N$ under the assumption that $p$ is a
  $k-j$-abundance distribution. Then $\E\left(\widehat{N}^{k-j}
  \right) = N(1 - \nabla^{k-j} p_{0})$. As
\begin{equation*}
 \nabla^{k-j} p_{0}  =  \nabla^{k}p_{0} + \sum_{h=0}^{j-1}
 \nabla^{k-j-h}p_{1} \mbox{ if } 1 \leq j \leq k-1 ,
\end{equation*}
$\nabla^{k} p_{0} =0$ and $\nabla^{k-j-h}p_{1} >0$ (recall that
$k$-monotone distributions are strictly $(k-j)$-monotone), we get that
$N$ is under-estimated. 
\end{rmk}
\begin{rmk}\label{Nkplusj.rmk}
If we estimate $N$ under the assumption that $p$ is a
  $(k+j)$-abundance distribution, then $\E\left(\widehat{N}^{k+j}
  \right) = N(1 - \nabla^{k+j} p_{0})$ where
\begin{equation*}
 \nabla^{k+j} p_{0}  =  - \sum_{h=1}^{j}
 \nabla^{k+j-h}p_{1} \mbox{ if } j \geq 1. 
\end{equation*}
In that case the estimator of $N$ is biased. If $j=1$, 
$\nabla^{k+j} p_{0} =  -  \nabla^{k}p_{1}$ which is negative or null,
and $N$ is over-estimated.
\end{rmk}

\subsection{Estimator based  of  the constrained least-squares
  estimator of  $p^{+}$}

The  empirical estimator  may be non $k$-monotone whereas under our
assumptions, $p^{+}=(p^{+}_{1}, p^{+}_{2}, \ldots)$ is a $k$-monotone
density. Hence, in addition to the empirical estimator $f=(f_1, f_2, \ldots)$, we
consider an estimator that takes into account the  constraint of
$k$-monotonicity. Precisely, we consider the constrained least-squares
estimator  $\widetilde{p^{+}}$  of $p^{+}$ defined as follows: 
\begin{equation}
\widetilde{p^{+}} = \mbox{arg} \min  \left\{ 
\sum_{j \geq 1} (q_j -  f_j)^{2},  
q = (q_{1}, q_2, \ldots), q  \mbox{ a } k\mbox{-monotone distribution} \right\}.
\label{ptilde.eq}
\end{equation}
    
Existence and uniqueness of $\widetilde{p^{+}}$ was studied
by~\cite{giguelay2017estimation}. Note that this reference considers
$k$-monotone distributions on $\N$ whereas we are interested here in
$k$-monotone distributions on $\N\backslash\{0\}$, but considering the
shifted distribution $p^{+}_{j+1}$ for $j\geq 0$, which is
$k$-monotone on $\mathbb{N}$, and the corresponding shifted estimators
$f_{j+1}$ and $\widetilde{p^{+}}_{j+1}$ allows to put our framework into
that of~\cite{giguelay2017estimation}, including the computation of the
estimator. In that paper the author  gives a characterization of
the estimator based on the
decomposition of $k$-monotone distributions as  mixtures of spline
functions. She showes that the least-squares estimator under the
constraint of $k$-monotonicity is closer (with respect to the  the
$\ell^{2}$-loss) to any $k$-monotone
distribution than the empirical distribution is.  Therefore, one could
expect that if $p^{+}$ is $k$-monotone, 
$\widetilde{p^{+}}^{k}$ will give better results, at least from the point of
view of the $\ell^{2}$-loss, than the empirical
distribution $f$. Moreover, the author 
implements the estimator using an exact iterative algorithm inspired
by the Support Reduction Algorithm described
in~\cite{groeneboom2008support} and  discusses
a practical stopping criterion.

Finally it remains to estimate $N$ by 
\begin{equation}
 \widetilde{N}^{k} = D \left(1 - \sum_{h=1}^{k}(-1)^{h} C_{k}^{h}
   \widetilde{p^{+}}^{k}_{h} \right).
\label{NTild.eq}
\end{equation}

\subsection{Estimating the degree of monotonicity of $p^{+}$}

For a given integer $k$, assuming that the distribution $p$ is a
$k$-abundance distribution, we propose two estimators of $N$,
$\widehat{N}^{k}$, see~\eref{NEmp.eq}, and $\widetilde{N}^{k}$, see~\eref{NTild.eq}. In practical cases, we do
not know the degree of monotonicity of $p$. Because we observe $X_{1},
\ldots, X_{D}$ with distribution $p^{+}$, we propose to estimate the
degree of monotonicity of $p^{+}$, 
using the method described in Section~\ref{Estk.st}.  Actually the
degrees of monotonicity of $p$ and $p^{+}$ are not necessarily
equal:  we know  that if $p$ is $k$-monotone then $p^{+}$ is at least
$k$-monotone, but to relate the degree of monotony of $p^{+}$ to the
one of $p$, we need an additional assumption on $p$. Precisely we
assume that  $p$
is a $k$-monotone abundance distribution, $p$ is not $(k+1)$-monotone
and $\nabla^{k+1}_{j_{0}} < 0$ for some $j_{0} \geq 1$. 
For example, the distributions $p^{h}=\cP(\lambda^{h}), h\leq 1$ defined at
Section~\ref{PoissonDist.st} and Table~\ref{lambdaValues.tab}, satisfy
\begin{eqnarray*}
\nabla^{h} p_{0} & \approx & 0 \\
\nabla^{h} p_{j} & > & 0 \mbox{ for all } j \geq 1 \\
\nabla^{h+1} p_{0} & < & 0  \\
\nabla^{h+1} p_{j} & \geq & 0  \mbox{ for all } j \geq 1.
\end{eqnarray*}
It comes that they do not satisfy the assumptions allowing to deduce
the degree of monotonicity of $p$ from the one of  $p^{+}$.

To sum up, we propose a procedure in two steps: at the first step we
estimate the degree of monotonicity of $p^{+}$ using the procedure
described at Section~\ref{Estk.st}. Let us denote by $\widehat{k}$ this
estimator. At the second step, we calculate
$\widehat{N}=\widehat{N}^{\widehat{k}}$ and $\widetilde{N}=\widetilde{N}^{\widehat{k}}$. We  assess
the performances of this procedure by simulation.

\subsection{Simulation experiment}

We construct the distributions $p$ as follows: we choose a distribution
$p^{+} = (p^{+}_{1}, p^{+}_{2}, \ldots)$ such that $p^{+}$ is
$k$-monotone but not $(k+1)$-monotone on the set of
integers greater than 1. Then we calculate $p$ such that $p_{0}$ satisfies
Equation~\eref{theta.eq}, and for all $j\geq 1$, $p_{j} =  (1-p_{0}) p^{+}_{j}$.

Given $N$ and $p^{+}$, a simulation consists in two steps: first we
draw one realization 
of $D$ distributed as a $\cB(N, 1-p_{0})$, then we draw $D$
realizations $X_{1}, \ldots, X_{D}$ distributed as $p^{+}$. From this
simulated sample, we estimate $p^{+}$ either by the empirical
dstribution $f$ or by the least-squares estimator under the constraint
of $k$-monotonicity, see Equation~\eref{ptilde.eq}.

We choose three values of $N$, $N=1000, 5000, 30000$, and 
five distributions $p^{+}$, denoted $p^{+, h}$, such that $p^{+,
  h}_{j} = p^{h}_{j-1}$ for $j \geq 1$ and for $h = \left\{1, \ldots, 5\right\}$
(see Section~\ref{PoissonDist.st} and Table~\ref{lambdaValues.tab} for
the definition of $p^{h}$). 

\subsubsection{Comparison of the estimators $\widehat{N}^{k}$ and $\widetilde{N}^{k}$}

\medskip

The calculation of $\widetilde{N}^{k}$ lies on the least-squares
estimator of $p^{+}$ under the constraint of $k$-monotonicity. For
$k=\left\{2, 3, 4\right\}$ the algorithm for estimating $p^{+}$
is available in the R-package \verb+pkmon+  on the Comprehensive R Archive
Network\footnote{https://CRAN.R-project.org/package=pkmon}. For 
$k=1$, we used the algorithm developped
by~\cite{reboul1998estimation}.

For each simulation we calculate
$\widehat{k}$ for  $k_{\max}=4$ using Procedure {\bf P1},  $\widehat{N}^{k}$  and
$\widetilde{N}^{k}$ for each $k=\left\{1, \ldots, 4\right\}$, 
and $\widehat{N} = \widehat{N}^{\widehat{k}}$  and $\widetilde{N}=\widetilde{N}^{\widehat{k}}$. We
report their expectation and prediction error estimated on the basis
of $S=500$ simulations. Precisely, if $\widehat{N}^{k}_{s}$ is the
estimation of $N$ at simulation $s$, we calculate
\begin{equation*}
\mathrm{PE}(\widehat{N}^{k}) = \frac{1}{S} \sum_{s} (\widehat{N}^{k}_{s} -N)^{2}.
\end{equation*}
We report in Table~\ref{comparEstimN.tb} $\widehat{N}^{k}_{\bullet}$, the mean of the
$\widehat{N}^{k}_{s}$'s and $100 \sqrt{\mathrm{PE}(\widehat{N}^{k})}/N$, as well as
$\widetilde{N}^{k}_{\bullet}$ and $100
\sqrt{\mathrm{PE}(\widetilde{N}^{k})}/N$. The bold values correspond
to the cases where the 
  estimation of $N$ is carried out assuming that the degree of
  monotonicity of the truncated distribution is known: $k=h$. The exponent
  denotes how many simulations failed to give the result. This may
  happen in the following situations:
\\
- The estimator
  $\widehat{N}^{k}$ can be calculated only if $ \sum_{j=1}^{k} (-1)^{j}
  C_{k}^{j}S_{j}$ is positive (see Remark~\ref{NsupD.rmk}). For example when $N=1000$, $k=3$, $h=1$ this
  condition was not satisfied in 11 simulations over 500. If $k=4$, there is no
  result because the condition was not satisfied in more than 1
  simulation over 2. Note that this condition is always satisfied for $ \sum_{j=1}^{k} (-1)^{h}
  C_{k}^{j}\widetilde{p^{+}_{j}}^{k}$ because $\widetilde{p^{+}}$ is $k$-monotone.
\\
- The algorithm for calculationg
  $\widetilde{N}^{k}$ may fail to converge for some simulation. For
  example when $N=1000$, $k=4$, $h=3$ or $h=4$ this happened 10
  times.
\\
-The estimator $\widehat{k}$ may equal 0, in particular when
  $h=1$: this is expected  in about $5\%$ of the simulations (the
  aymptotic level of the testing procedure). When $N=1000$, and $h=1$
  this happenned in 14 simulations.

\medskip 

Let us now comment the results. 
\begin{itemize}
\item If the degree of monotonicity of
$p^{+}$ is known (cases in bold where $k=h$), the estimators behave similarly with a small
advantage for $\widetilde{N}^{k}$ whose prediction error is
smaller. As expected $\widehat{N}^{k}$ is unbiased which is not the
case of $\widetilde{N}^{k}$. However 
$\widetilde{N}^{k}$ has a  smaller  variance than 
$\widetilde{N}^{k}$, smaller enough to have a smaller prediction
error. 
\item   When $k$ is strictly smaller than $h$, then $\widehat{N}^{k}$ and $\widetilde{N}^{k}$
are nearly always equal. This comes from the fact that   $p^{h}$
is strictly $k$-monotone. Therefore, because $N$ is large enough, the
empirical distribution is  nearly always $k$-monotone, Let us note that if the empirical distribution is $k$-monotone,
then the least-squares estimator under the constraint of
$k$-monotonicity is exactly equal to the empirical distribution. 
\item When $k$ is strictly greater than $h$, then $\widehat{N}^{k}$
  tends to underestimate $N$ while $\widetilde{N}^{k}$ tends to
  overestimate it. This behaviour of $\widetilde{N}^{k}$ was expected,
  see Remark~\ref{Nkplusj.rmk}. 
\item When $k=\widehat{k}$,  let us consider the cases where $h \leq
  3$. Indeed, as $k_{\max} = 4$, we know that $\widehat{k}$ is nearly
  always equal to $k_{\max}$ when $h \geq k_{\max}$. When $h=1, 2, 3$,
  taking $k=\widehat{k}$
  for estimating $N$ leads to increase the 
  prediction error with respect to the case $k=h$. This tendancy is
  more pronounced for $\widetilde{N}$.
\end{itemize}

\begin{table}
\begin{center}

$N=1000$

\bigskip
\begin{footnotesize}
\begin{tabular}{c|ccccc|ccccc}
& \multicolumn{5}{c|}{$\widehat{N}^{k}$}   &  \multicolumn{5}{|c}{$\widetilde{N}^{k}$} \\\cline{2-11}
      &$h=5$&$h=4$&$h=3$& $h=2$& $h= 1$    & $h=5$   & $h=4$   & $h=3$   & $h=2$   & $h= 1$ \\\hline
$k=1$ & 573 & 653 & 749 & 870  & {\bf 999}       & 573& 653 & 749 & 870 &{\bf 1003} \\
      & 43& 35& 26& 13 & 2.3      & 43& 35&  25&  13& 2.1\\\hline
$k=2$ & 756 & 840 & 921 &{\bf 997}  &1000       & 756& 840& 922 &{\bf 1007}&1111 \\
      & 25& 17& 8.9& 4.2 & 3.9      & 25& 17&  8.9&3.7& 11\\\hline
$k=3$ & 882 & 952 & {\bf 996} &996  &$872^{(11)}$& 882 & 953 &  ${\mathbf{ 1009^{(1)}}}$ &${{ 1073^{(4)}}}$ & 1151 \\
      & 13& 8.0& 6.4& 6.9 &  14     &13& 7.9& 5.4& 8.3& 15 \\\hline
$k=4$ & 958 &{\bf 1003} & 994 &908   &           & 963 & ${\mathbf{ 1020^{(10)}}}$ & $1058^{(10)}$ &$1112^{(2)}$ & 1169 \\
      & 9.5& 9.4&9.9& 14&           &  8.7& 8.1&  8.3& 12&  17\\\hline
$k=\widehat{k}$&958&1004&1007&995&$988^{(14)}$&962 &$1016^{(10)}$ &$1041^{(10)}$ &$1049^{(1)}$&$1012^{(14)}$ \\
               & 9.4& 9.2& 8.5& 6.6&2.6& 8.8&8.4& 8.6& 9.4& 4.4
\end{tabular}
\end{footnotesize}

\bigskip

 $N=5000$
\bigskip

\begin{footnotesize}
\begin{tabular}{c|ccccc|ccccc}
& \multicolumn{5}{c|}{$\widehat{N}^{k}$}   &  \multicolumn{5}{|c}{$\widetilde{N}^{k}$} \\\cline{2-11}
      &$h=5$&$h=4$&$h=3$& $h=2$& $h= 1$        & $h=5$   & $h=4$   & $h=3$   & $h=2$   & $h= 1$ \\\hline
$k=1$ & 2879 & 3262 & 3744 & 4358  & {\bf 4998}& 2879& 3262 & 3744 & 4357 &{\bf 5007} \\
      & 42 & 35 & 25 & 13  & 0.98    & 42& 35&  25&  13& 0.9\\\hline
$k=2$ & 3801 & 4192 & 4613 &{\bf 5003}&5001    & 3800& 4192& 4613 &{\bf 5023}& 5557 \\
      & 24& 16& 7.9& 1.9 & 1.7       & 24& 16&  7.9&1.7& 11\\\hline
$k=3$ & 4436 & 4751 & {\bf 4990} & 5001  & 4332& 4436 & 4751 & {\bf 5021} &$5367^{(6)}$ & 5760 \\
      & 12& 5.7& 3.1& 3.1 &  13.7    &12& 5.7& 2.6& 7.6& 15 \\\hline
$k=4$ & 4826 &{\bf 5004} & 4987 & 4568   &     & 4827 & ${\mathbf{ 5042^{(9)}}}$ & $5254^{(17)}$ &5570 & 5842 \\
      & 5.1& 4.1     &4.9 & 10   &     &  5.1& 3.5&  5.7& 12&  17\\\hline
$k=\widehat{k}$&4826&5002&5011&4977&$4998^{(9)}$&4827 &$5032^{(8)}$ &$5063^{(1)}$ &4994&$5007^{(9)}$ \\
               &5.2& 4.2&4.3&3.5&0.97& 5.1&3.9& 4.8& 3.6& 0.92
\end{tabular}
\end{footnotesize}

\bigskip

 $N=30000$

\bigskip

\begin{footnotesize}
\begin{tabular}{c|ccccc|ccccc}
& \multicolumn{5}{c|}{$\widehat{N}^{k}$}   &  \multicolumn{5}{|c}{$\widetilde{N}^{k}$} \\\cline{2-11}
      &$h=5$&$h=4$&$h=3$& $h=2$& $h= 1$        & $h=5$   & $h=4$   & $h=3$   & $h=2$   & $h= 1$ \\\hline
$k=1$ &17270&19560&22499&26128&{\bf 29997}     &17270&19560&22499&26128&{\bf 30025} \\
      & 42 & 35 & 25 & 13  & 0.42              & 42 & 34&  25&  13& 0.40\\\hline
$k=2$ &22795&25119&27720&{\bf 29992}&29982     &22796&25119&27720&{\bf 30045}&33364 \\
      & 24  & 16& 7.6& 0.82 & 0.73             & 24& 16&  7.6& 0.71 & 11\\\hline
$k=3$ &26607&28453&{\bf 29994} &29980&25923    &26607&28453&{\bf 30060} &$32166^{(16)}$ &34583 \\
      & 11& 5.3& 1.1& 1.3 & 13                 &11& 5.3& 0.98& 7.2& 15 \\\hline
$k=4$ &28944&{\bf 29946} &29988&27381&         &28944& ${\mathbf{ 30034^{(12)}}}$&$31484^{(17)}$&33421&35063 \\
      & 3.8&1.7     &1.7 & 8.9   &             & 3.8& 1.4& 5.0& 11&  17\\\hline
$k=\widehat{k}$&28944&28898&29904&29796&$30001^{(16)}$&28944&29965&29954&29836&$30026^{(16)}$ \\
               &3.8& 2.1&2.1&3.2&0.42          & 3.8&1.9& 2.1& 3.2& 0.4
\end{tabular}
\end{footnotesize}

\end{center}
\caption{\label{comparEstimN.tb} On the left hand side: values of
  $\widehat{N}^{k}_{\bullet}$ and $100
  \sqrt{\mathrm{PE}(\widehat{N}^{k})}/N$ for $k=1, \ldots, 4$ and
  $k=\widehat{k}$. On the right hand side: the same for $\widetilde{N}^{k}$.}
\end{table}

\subsubsection{Effect of $N$ and $k_{\max}$ on $\widehat{N}$.}

For several values of $k_{\max}$, $k_{\max} \in \left\{4, 6, 10\right\}$,
we estimate the expectation and prediction error of
$\widehat{N}=\widehat{N}^{\widehat{k}}$  over 500 simulations. The
results are given at Table~\ref{EffetkmaxEstimN.tb}. As in Table~\ref{comparEstimN.tb}, the exponent denotes
how many simulation failed to give the result.  When $N=1000$,
we know (see Table~\ref{SimulkChapPoissonP1.tab}) that $\widehat{k}$ over-estimates $h$: for example, when $h=3$
we get $\widehat{k}=3$ in 180 simulations while we get $\widehat{k}
\geq 6$  in  210
simulations. This leads  to increase the variability of
$\widehat{N}$. Moreover,  when $k_{\max}$ increases, the calcuation of
$\widehat{N}$ becomes impossible (in one simulation over 5 for $h=3$
and $k_{\max}=10$). 

As expected, when $N$ increases, the prediction error of
$\widehat{N}$ decreases. 
If $k_{\max}$ is chosen smaller than $h$, then
$\widehat{N}$ under-estimates $N$. If $k_{\max}$ is greater than $h$,
 then the loss in terms of prediction error
between $\widehat{N}^{h}$ and $\widehat{N}$ decreases with $N$. For example
if $k_{\max}=10$, $N=30000$ and $h=4$, the prediction error for
$\widehat{N}^{4}$ equals 1.7 (see Table~\ref{comparEstimN.tb}) while
it equals 2.3 for $\widehat{N}$.

\begin{table}
\begin{center}

\bigskip

$N= 1000$

\bigskip

\begin{tabular}{l|ccccc}
& $h= 5$ & $h= 4$ & $h= 3$ & $h= 2$ & $h= 1$ \\ \hline
&  968 & 999    & 1015   &  989  & $998^{(20)}$ \\
$k_{\max}=4$& 9.4  &  8.9  & 9.2 & 6.9   & 2.7        \\ 
&4 & 4 & 4 & 2 & 1
\\ 
\hline
&  1013 & 1004    &  972   &  948   & $1002^{(18)}$ \\
$k_{\max}=6$& 15  &  14  & 14   & 12   & 3.1         \\ 
& 6 & 6 & 6 & 2 & 1
\\ 
\hline
& $932^{(28)}$ & $942^{(66)}$ & $1024^{(101)}$ & $1159^{(48)}$ & $1035^{(13)}$ \\
$k_{\max}=10$ & 37   & 67   & 108 & 45 & 24        \\ 
& 10 & 10 & 10 & 2 & 1 \\ 
\end{tabular}

\bigskip

$N= 5000$

\bigskip

\begin{tabular}{l|ccccc}
& $h= 5$ & $h= 4$ & $h= 3$ & $h= 2$ & $h= 1$ \\ \hline
& 4816 & 5002   &  5001  &  49980  & $5000^{(18)}$ \\
$k_{\max}=4$&5.3  &  4.2  & 4.2   & 3.2   & 1.0         \\ 
&4 & 4 & 3 & 2 & 1 \\ 
\hline
&  5025 & 5015   & 4979   &  4957  & $5000^{(17)}$ \\
$k_{\max}=6$&6.8 & 5.2   & 4.2  & 3.9   & 1.0        \\ 
&6&5&3&2&1
\\ 
\hline
&4692  & 4650 & 4931 & 4985 & $5010^{(12)}$ \\
$k_{\max}=10$& 15  & 14  &9.3 & 3.9 & 1.0        \\ 
& 4 & 4 & 3 & 2 & 1 \\ 
\end{tabular}

\bigskip

$N= 30000$

\bigskip

\begin{tabular}{l|ccccc}
& $h= 5$ & $h= 4$ & $h= 3$ & $h= 2$ & $h= 1$ \\ \hline
& 28932  & 29952 & 29900 & 29833 & $30000^{(18)}$ \\
$k_{\max}=4$& 3.9  & 2.1  & 2.2 & 2.9 & 0.43        \\ 
& 4 & 4 & 3 & 2 & 1 \\ 
\hline
&  30075 & 29980    & 29904   &  29874  & $30009^{(14)}$ \\
$k_{\max}=6$& 2.8  &  2.1  & 2.2  & 2.6   & 0.41         \\ 
&5 & 4 & 3 & 2 & 1
\\ 
\hline
& 29989 & 29932 & 29848 & 29843 & $30000^{(18)}$ \\
$k_{\max}=10$ & 2.8  & 2.3   & 2.3 & 2.9 & 0.42         \\ 
& 5 & 4 & 3 & 2 & 1 \\ 
\end{tabular}
\end{center}
\caption{\label{EffetkmaxEstimN.tb} For each value of $k_{\max}$, the
  first line reports, for each value of $h$, 
  $\widehat{N}^{\widehat{k}}_{\bullet}$, the second line  $100
  \sqrt{\mathrm{PE}(\widehat{N}^{\widehat{k}})}/N$ and
  the third line reports the median of $\widehat{k}$.}
\end{table}

\section{\label{realData.st} Application to real data sets}

Most real observed abundance distributions are at least decreasing and
appear to be $k$-monotone for some $k \geq 2$. Several examples were
already studied when considering convexity~\cite{capture, durot2015nonparametric}. Let
us consider three examples in order to illustrate how our procedure
applies when we aim at estimating the total number of classes taking
into account the  hollowed shape of the abundance distribution. 
\begin{enumerate}
\item Data from Hser~\cite{Hser01}, given Table 1.3
in~\cite{captureIntro}, reporting the episode count per
drug user in Los Angeles 1989, are used for estimating the size of a
population of illicit drug users.
\item The famous data set reporting the frequencies of word types used
  by Shakespeare (see~\cite{Spevack1968complete} and Table 1.8 in~\cite{captureIntro}),
  allows to estimate  {\it how many words did Shakespeare
know but not use}~(Efron and Thisted~\cite{efron1976estimating}).
\item A metagenomics data set~\cite{tap2009towards} was analysed by Li-Thiao-Té et al.~\cite{li2012bayesian} to estimate the total number of microbial strains in
the human gut microbiome.
\end{enumerate}

For the two last data sets, the maximum of the support of the
empirical abundance distribution is very large: five words were seen 100 times,
one strain were seen 564 times. Indeed, the tail of the distribution
does not contribute to estimate the behaviour of the beginning of
the distribution. However considering a very large number of variates
in the test statistics may affect the power of the test by increasing
$-\widehat{q}_{\alpha}^{k}$  in procedure {\bf P1} and 
$-\widehat{u}^{k}_{\alpha}$ in procedure {\bf P2}. Therefore we carried
out the testing procedure replacing  $\widehat{\tau}$ in the definition
of $\widehat{\T}^{k}$ and $\widehat{\U}^{k}$ by the minimum of some
fixed integer $l$ and $\widehat{\tau}$. The results are given with
$l=20$. For these two data sets, it appears that the results does not
change with the value of $l$.

\begin{figure}
\includegraphics[height=.45\textwidth, width=.45\textwidth,angle=0]{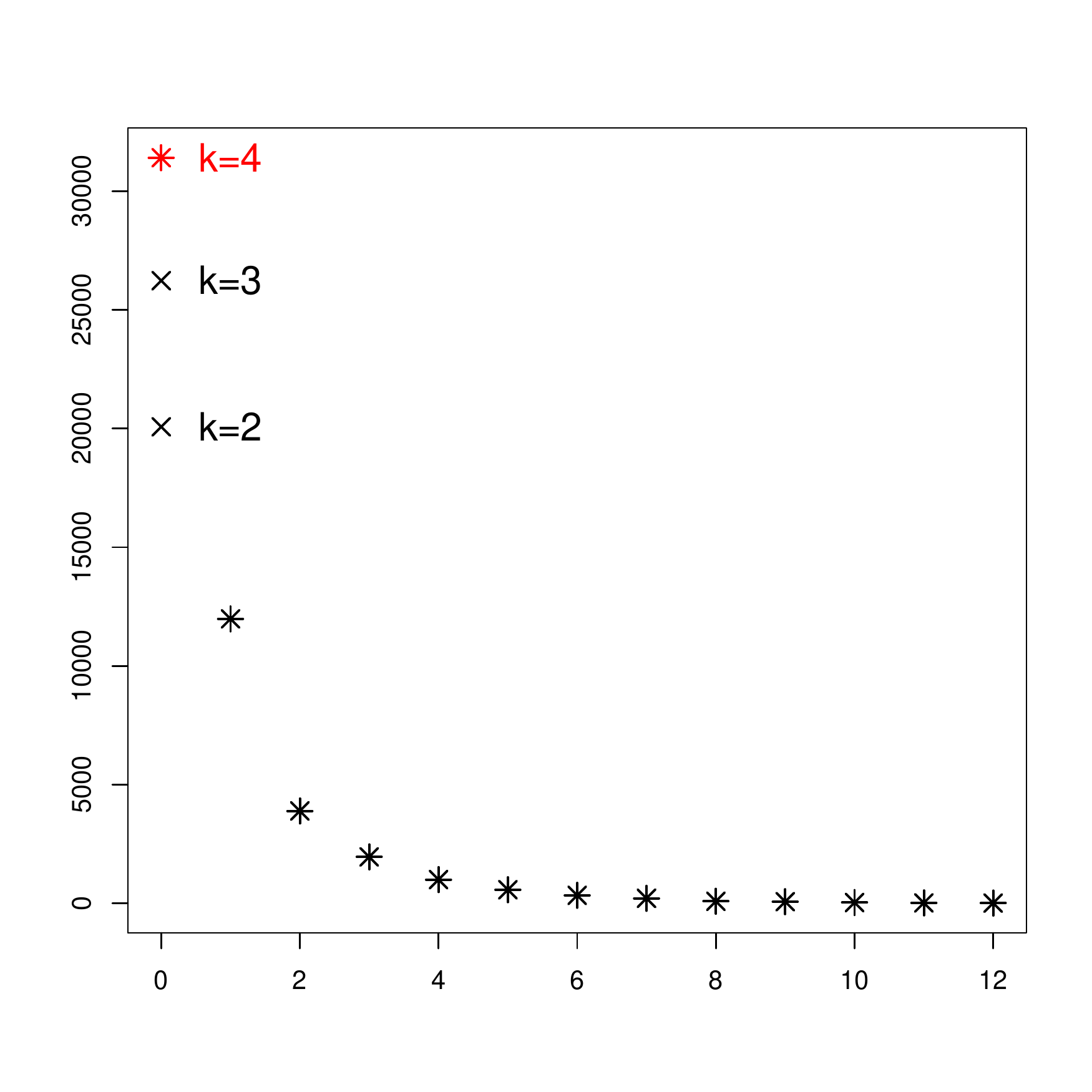}
\includegraphics[height=.45\textwidth, width=.45\textwidth,angle=0]{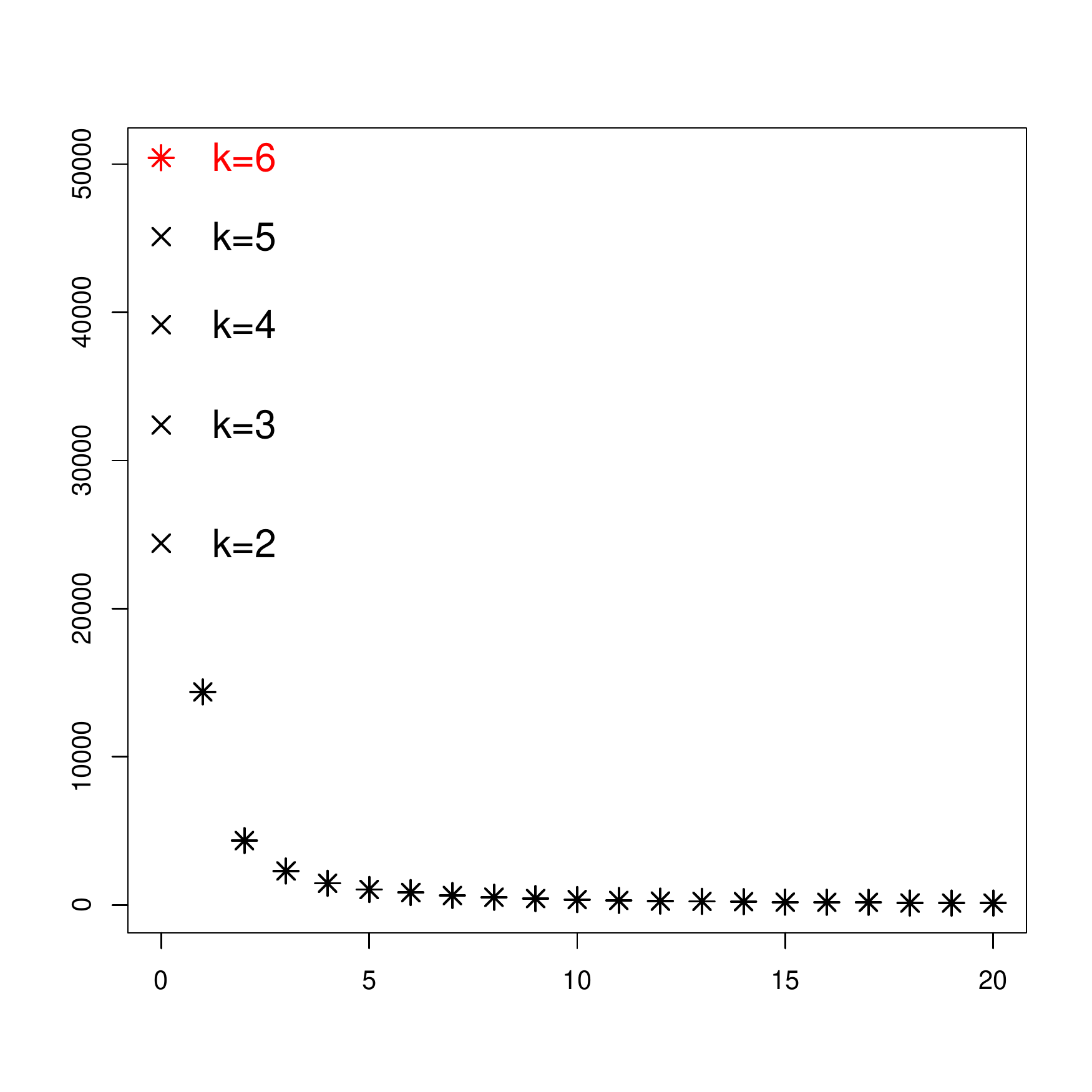}
\includegraphics[height=.45\textwidth, width=.45\textwidth,angle=0]{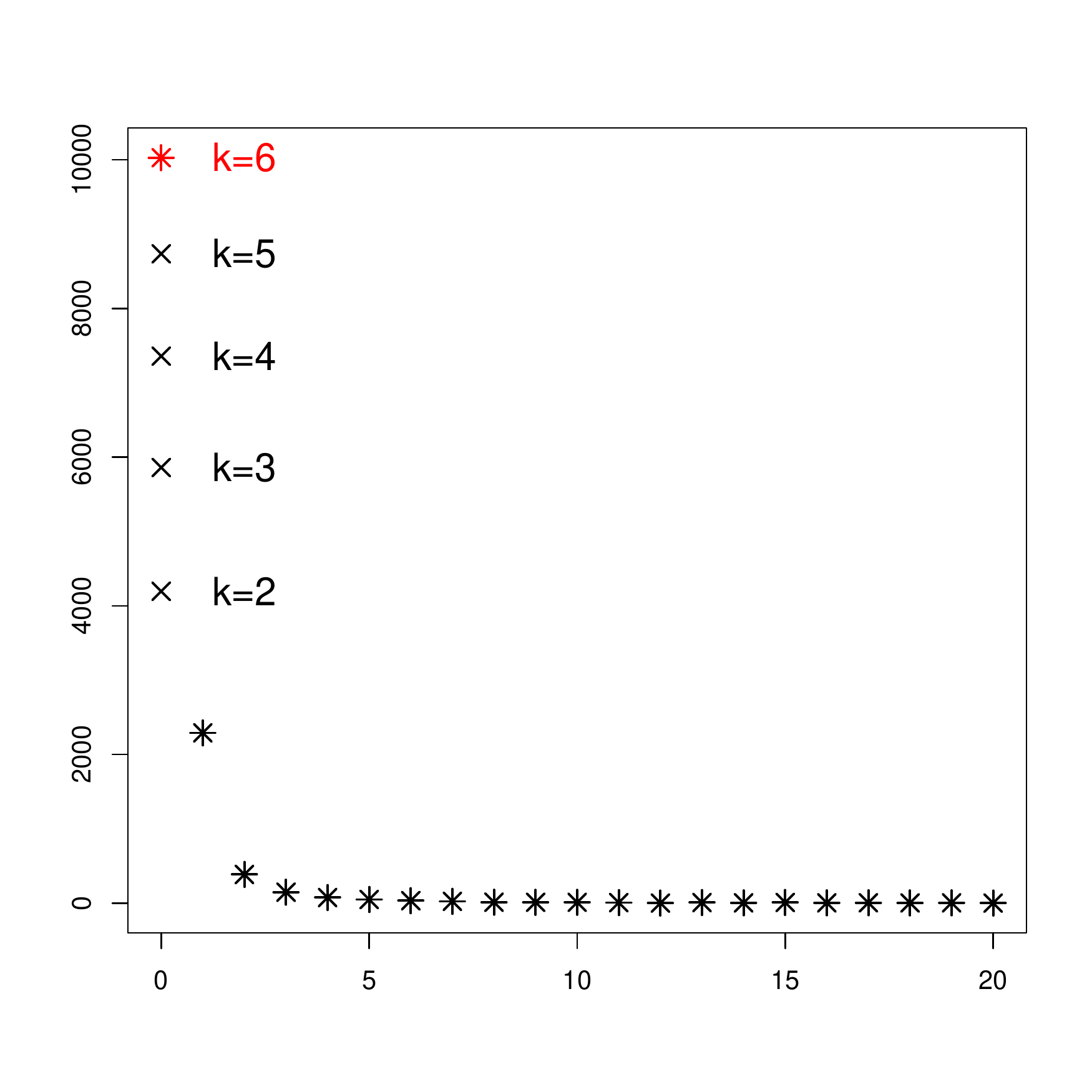}

\caption{\label{TestEstimN.fig} For each data set, graphics of the
  observed frequencies on the interval [1, 20]. For each $2 \leq k
  \leq \widehat{k}$, estimated value of the number of missing classes
  or species.}
\end{figure}

For each data set we test the hypothesis $H^{k}$ for $1 \leq k \leq
k_{\max}$ with $k_{\max}=6$, and calculate the estimated number of classes as well as its
estimated standard-error. The results are given in
Table~\ref{TestEstimN.tb} and Figure~\ref{TestEstimN.fig}.

\begin{table}
\begin{center}

Size of a population of drug users : $D=20198$ episodes counted

\medskip 

\begin{tabular}{c|cc|rr}
 & Test {\bf P1} & Test {\bf P2} &$ \widehat{N}^{k}$ &
 $\widehat{s}^{k}$ \\
$k=1$ & accept & accept & 32180 & 154 \\
$k=2$ & accept & accept & 40269 & 268 \\
$k=3$ & accept & accept & 46424 & 414 \\
{\color{red} $k=4$} & accept & accept & {\color{red} 51602} & {\color{red} 629} \\
$k=5$ & accept & reject & 56333 & 973 \\
$k=6$ & reject & reject & 60955 & 1542 \\
\end{tabular}

\medskip

Number of  words Shakespeare knew : $D= 30709$ words used

\begin{tabular}{c|cc|rr}
 & Test {\bf P1} & Test {\bf P2} &$ \widehat{N}^{k}$ &
 $\widehat{s}^{k}$ \\
$k=1$ & accept & accept & 45085 & 170 \\
$k=2$ & accept & accept & 55118 & 298 \\
$k=3$ & accept & accept & 63100 & 451 \\
$k=4$ & accept & accept & 69860 & 681 \\
$k=5$ & accept & accept & 75807 & 1051 \\
{\color{red} $k=6$} & accept & accept & {\color{red} 81136} & {\color{red} 1682} \\
\end{tabular}

\medskip

Number of microbial strains in the human gut microbiome : $D= 3180$
strains seen

\medskip

\begin{tabular}{c|cc|rr}
 & Test {\bf P1} & Test {\bf P2} &$ \widehat{N}^{k}$ &
 $\widehat{s}^{k}$ \\
$k=1$ & accept & accept & 5471 & 68 \\
$k=2$ & accept & accept & 7375 & 117 \\
$k=3$ & accept & accept & 9040 & 173 \\
$k=4$ & accept & accept & 10538 & 245 \\
$k=5$ & accept & accept & 11915 & 348 \\
{\color{red} $k=6$} & accept & accept & {\color{red} 13207} & {\color{red} 508} \\
\end{tabular}
\end{center}
\caption{\label{TestEstimN.tb} For each data set, decision of the test
 of the hypothesis $H^{k}$, estimation of $N$ and of the
 standard-error of $\widehat{N}^{k}$, for $1 \leq k \leq 6$. The line
 in red corresponds to the estimated value of $k$. }
\end{table}

For the first example we choose $\widehat{k}=4$ using {\bf P1} and
$\widehat{k}=5$ using 
{\bf P2}, while for the two last data sets, we choose
$\widehat{k}=k_{\max}$. This choice may be explained by the followed shape of
the empirical distributions together with the difficulty of rejecting
$H^{k}$  for large $k$. 

The number of Shakespeare's unused words was estimated to be at least
equal to 35000 by~\cite{efron1976estimating}.  Using our procedure
with $k_{\max}=6$, we get approximatively 50000 words. In that example
$D$ is large enough to protect us against lack of power for testing
$H^{k}$, at least for $k \leq 4$. Therefore we are confident that
$k=6$ is a reasonable choice.

Concerning the number of strains, the estimation given by the Chao1
procedure~\cite{chao1984nonparametric}, $\widehat{N} = D + f_{1}^{2}/2
f_{2}$, equals 9940, while the estimation given
by~\cite{li2012bayesian} is 25700 with a $95\%$ confidence interval
equals to $[19421, 36355]$. Choosing $k=6$ we get $\widehat{N} =
13207$. Let us see (Table~\ref{TestOTU.tb}) what happens if $k$
increases: for any $7 \leq k \leq 13$ the hypothesis $H^{k}$ is not
rejected. If $k=12$ we get  $\widehat{N} = 23561$ which is close to
the value proposed by Li-Thiao-Té et al.~\cite{li2012bayesian}. Nevertheless, the
estimated standard-error 
$\widehat{s}^{k}$ increases drastically with $k$,  making the result
useless for large $k$.

\begin{table}
\begin{center}
Number of microbial strains in the human gut microbiome : $D= 3180$
strains seen

\medskip

\begin{tabular}{cccccccc}
& $k=7$ & $k=8$ &$k=9$ &$k=10$&$k=11$ & $k=12$&$k=13$ \\
$\widehat{N}^{k}$ &14447& 15675 & 16962 & 18458& 20469& 23561& 28695\\
$\widehat{s}^{k}$ & 770 & 1218& 2004& 3401& 5900& 10378& 18411
\end{tabular}
\end{center}
\caption{\label{TestOTU.tb}  Estimation of $N$ and of the
 standard-error of $\widehat{N}^{k}$, for $7 \leq k \leq 13$. }
\end{table}

\section{Conclusion\label{Conclusion-test-st}}

We proposed two testing procedures and their boostrap versions to
test the null hypothesis that a discrete distribution is $k$-monotone
against that it is not, without any parametric assumption on the true
underlying distribution. We state the theoretical asymptotic
properties of the procedures and carry out a large simulation study in
order to assess their performances for finite sample cases. The
simulation shows that the tests may present a power fault and require
large values of the sample size $d$, in particular when $k$ is large. We compare this
non-parametric setting with a parametric procedure for Poisson
distribution, when the problem is to test the null hypothesis that the
distribution is at least $k$-monotone against the alternative that it is
$(k-1)$-monotone but not $k$-monotone. We conclude that the efficiency
(the sample size required for the test to achieve a given power)
of the non-parametric procedure is much more affected for large values
of $k$ than the parametric procedure is. The comparison with the
procedure of~\cite{balabdaoui2017testing} based on the $\ell^{2}$
distance between the constraint least-squares estimator and the
empirical estimator for testing convexity suggests potential
improvements.

From these testing procedures we propose a 
method  to infer the degree of
$k$-monotonicity of a discrete distribution, assuming that $k$ is
smaller than some $k_{\max}$. To
our knowledge this is the first method for estimating the degree of
$k$ monotonicity of discrete distribution for which theoretical
guaranties  are established. A large simulation study shows  that
the performance of the estimator of $k$ depends strongly on the choice
of $k_{\max}$: large values of $k_{\max}$ need large sample sizes.

Finally we apply  this work to the estimation of the unknown number of
classes in a population. Defining a $k$-monotone abundance
distribution, the identifiability of the parameter to estimate is
ensured. A simulation study shows that the method can be applied
providing that the number of seen classes is large, especially as $k$
increases.

\section{Proofs\label{proofs-tests.st}}

\subsection{Proof of Theorem~\ref{TestkLevel.th}}

Let us first remark that for $d$ large enough, $\widehat{\tau}$ is almost
surely equal to $\tau$. This result comes from the application of the
Borel-Cantelli lemma, by noting that
\begin{equation*}
\sum_{d=1}^{\infty} P\left( \widehat{\tau} < \tau\right) =
\sum_{d=1}^{\infty}(1-p_{\tau})^{d} < + \infty.
\end{equation*}

In the following we will assume that $d$ is large enough to set
$\widehat{\tau}=\tau$.

\paragraph{Procedure P1} Let us begin with the testing procedure based on the statistic $\widehat{\T}^{k}$.
Because $p$ is $k$-monotone, 
\begin{eqnarray*}
 P\left( \sqrt{d} \min_{0\leq j \leq \tau-1}
     \nabla^{k}f_{j} \leq q\right) & \leq &
P\left( \sqrt{d} \min_{0\leq j \leq \tau-1}
     \left(\nabla^{k}f_{j} - \nabla^{k}p_{j}\right)\leq q\right)
\end{eqnarray*}

By the central limit theorem we know that the vector $\sqrt{d} A
(f-p)$ converges in distribution to a centered Gaussian vector
with covariance matrix $A \Gamma A^{T}$, where $\Gamma$  is the matrix
     with components $\Gamma_{j j'} = - p_{j} p_{j'} $
     if $j\neq j'$ and $\Gamma_{j j} =
    p_{j}(1-p_{j})$ for $0 \leq j \leq \tau-1$. Let $M$  be
    defined as the square-root of the matrix $A \Gamma A^{T}$, then 
\begin{eqnarray*}
 P\left( \sqrt{d} \min_{0\leq j \leq \tau-1}
     \left(\nabla^{k}f_{j} - \nabla^{k}p_{j}\right)\leq
     q\right) &=&
 P\left( \sqrt{d} \min_{0\leq j \leq \tau-1}
     \sum_{j'=0}^{\tau}A_{jj'} (f_{j'} - p_{j'})\leq
     q\right) \\
&=& P\left( \min_{0\leq j \leq \tau-1} \sum_{j'=0}^{\tau-1}
     M_{jj'}\Z_{j'} \leq q \right) + o(1).
\end{eqnarray*}
uniformly for all $q \in \R$, where the $\Z_{j'}, j'=0, \ldots \tau-1$ are independent centered
Gaussian variates.
\\

Because $\widehat{\Gamma}$ converges in probability to $\Gamma$ when $d$
tends to infinity, and thanks
to the continuity of the limiting distribution of $\sqrt{d} \min_{j}
     \left(\nabla^{k}f_{j} - \nabla^{k}p_{j}^{+}\right) $, we get that
\begin{equation*}
 P\left( \min_{0\leq j \leq \tau-1} \sum_{j'=0}^{\tau-1}
     M_{jj'}\Z_{j'} \leq \widehat{q}^{k}_{\alpha} \right) = P\left( \min_{0\leq j \leq \tau-1} \sum_{j'=0}^{\tau-1}
     \widehat{M}_{jj'}\Z_{j'} \leq \widehat{q}^{k}_{\alpha} \right)+ o(1) = \alpha + o(1).
\end{equation*}

\paragraph{}
Let us consider now the case where $p$ is strictly $k$-monotone and
let $C \geq 0 $ be such that $\min_{0\leq j \leq \tau-1} \nabla^{k}p_{j} \geq
C$. 

\begin{eqnarray*}
 P\left( \sqrt{d} \min_{0\leq j \leq \tau-1}
     \nabla^{k}f_{j} \leq q\right) & = &
P\left( \sqrt{d} \min_{0\leq j \leq \tau-1}
     \left(\nabla^{k}f_{j} - \nabla^{k}p_{j}\right)\leq q  - \sqrt{d} C
     \right) \\
 & = & P\left( \min_{0\leq j \leq \tau-1} \sum_{j'=0}^{\tau-1}
     M_{jj'}\Z_{j'} \leq q - \sqrt{d} C \right) + o(1).
\end{eqnarray*}

Let $q_{\alpha}^{k}$ be defined at Equation~\eref{AsThatk.eq} and
$\sigma^k$ be defined in Theorem~\ref{TestkLevel.th}. Applying the Cramer-Chernoff method to Gaussian
     variables (see for 
example~Massart, 2003, chapter 2), we get the following result: 
\begin{eqnarray}
\label{aux15juin}
 P\left( \min_{0\leq j \leq \tau-1} \sum_{j'=0}^{\tau-1}
     M_{jj'}\Z_{j'} \leq q^{k}_{\alpha} -  \sqrt{d} C \right)  \leq  \tau \exp\left(
     -\frac{(q^{k}_{\alpha} -  \sqrt{d} C)^{2}}{ 2 (\sigma^{k})^{2}}\right).
 \end{eqnarray}
Then we have: 
 \begin{eqnarray*}
 P\left( \min_{0\leq j \leq \tau-1} \sum_{j'=0}^{\tau-1}
     M_{jj'}\Z_{j'} \leq q^{k}_{\alpha} -  \sqrt{d} C \right) \leq\tau \exp\left(
     -\frac{d C^{2}}{ 2 (\sigma^{k})^{2}}\right)
     \leq  \beta
\end{eqnarray*}
as soon as $C \geq \sqrt{2/d} \sigma^{k} \sqrt{\log( (\tau)/\beta)}$.
\\

\paragraph{Procedure P2} Let us now consider the procedure based on
$\widehat{\cS}_{\alpha}^{k}$. The proof of the first part of the
theorem is similar  to the proof
for the procedure {\bf P1}. Let us consider the case where $p$ is
strictly $k$-monotone. If $\min_{j} \nabla^{k}_{j} \geq C$,
$\zeta^{k}_{j} = \sqrt{ A^{k T}_{j}\Gamma A^{k}_{j}}$ and and  $ \zeta^{k} = \max_{j}\zeta_{j}^{k}$,
\begin{eqnarray*}
\P\left( \widehat{\cS}^{k}_{\alpha} \leq 0\right) & = & 
\P\left( \min_{0\leq j \leq \tau-1}
 \left\{\sqrt{d} \nabla^{k}{f}_{j} - 
\nu_{u^{k}_{\alpha}} \zeta^{k}_{j}\right\}
\leq 0\right) \\
& \leq & 
\P\left( \min_{0\leq j \leq \tau-1}
 \left\{ A^{k T}_{j} \Gamma^{1/2} \Z - 
\nu_{u^{k}_{\alpha}} \zeta^{k}_{j}\right\}
\leq - \sqrt{d} C  \right) + o(1)
\end{eqnarray*}
Then:
\begin{eqnarray}
\label{aux15juin4}
\P\left( \widehat{\cS}^{k}_{\alpha} \leq 0\right) \leq  \tau \max_{j}\; \P\left( A^{k T}_{j} \Gamma^{1/2} \Z \leq
\nu_{u^{k}_{\alpha}} \zeta^{k}_{j} - \sqrt{d} C   \right) + o(1)  
\end{eqnarray}
Moreover,
\begin{eqnarray*}
\P\left( A^{k T}_{j} \Gamma^{1/2} \Z \leq
\nu_{u^{k}_{\alpha}} \zeta^{k}_{j} - \sqrt{d} C   \right) \leq \exp\left(- \frac{\left(\nu_{u^{k}_{\alpha}} \zeta^{k}_{j} - \sqrt{d} C  \right)^2}{2(\zeta^{k}_{j})^{2}}\right).
\end{eqnarray*}
Then we have:
\begin{equation*}
\P\left( \widehat{\cS}^{k}_{\alpha} \leq 0\right)\leq 
\tau \max_{j}\;\exp\left(- \frac{ d C^2}{2(\zeta^{k}_{j})^{2}}\right)+o(1)
\leq  \beta+o(1)
\end{equation*}
as soon as 
\begin{equation*}
C \geq \frac{1}{\sqrt{d}}  
\left(\zeta^{k+1} \sqrt{2\log\frac{\tau}{\beta}}
 \right).
\end{equation*}

\subsection{Proof of Theorem~\ref{TestkPower.th}}
 
\paragraph{Procedure P1} 
 Let $q <0$ and $C$  such that $\nabla^{k+1}p_{j_{0}} \leq -C$,
\begin{eqnarray*}
 \P\left(  \widehat{\T}^{k+1} \geq
  q \vert D=d\right) & \leq &
\P\left( \sqrt{d} \nabla^{k+1}f_{j_{0}} \geq
  q\right) \\
 &\leq &
\P\left( \sqrt{d} (\nabla^{k+1}f_{j_{0}} -\nabla^{k+1}p_{j_{0}})\geq
  q - \sqrt{d} \nabla^{k+1}p_{j_{0}}\right) \\
 &\leq &
\P\left( \sqrt{d} (\nabla^{k+1}f_{j_{0}}
  -\nabla^{k+1}p_{j_{0}})\geq q+C \sqrt{d}\right)\\
 &\leq &
\P\left( \sqrt{d} A^{k+1 T}_{j_{0}} \Z \geq q+C
  \sqrt{d} \right) + o(1)
\end{eqnarray*}
Applying the classical Tchebychev inequality, we get
\begin{equation}
\label{aux15juin2}
 \P\left( \sqrt{d} A^{k+1 T}_{j_{0}} \Z \geq
  q^{k+1}_{\alpha} + C\sqrt{d} \right) \leq 
\exp\left(-\frac{(q^{k+1}_{\alpha} + C\sqrt{d})^{2}}{2 (\zeta^{k+1}_{j_{0}})^{2}}\right).
\end{equation}
Let us remark that applying Formula (\ref{aux15juin}) to the case where  $C=0$,  we get
\begin{equation*}
\alpha\leq  \tau \exp\left(
     -\frac{(q^{k+1}_{\alpha})^{2}}{ 2 (\sigma^{k+1})^{2}}\right)
\end{equation*}
then 
\begin{equation}
\label{maj.q.eq}
q_{\alpha}^{k+1}  \geq -\sigma^{k+1} \sqrt{2\log\frac{\tau}{\alpha}}.
\end{equation}
Considering inequalites given at Equations \eref{aux15juin2} and \eref{maj.q.eq}, we get that 
\begin{equation*}
 \P_{\bar{H}^{k+1}(C)}\left( \sqrt{d} A^{k+1 T}_{j_{0}} \Z \geq
  q^{k+1}_{\alpha} + C\sqrt{d} \right) \leq \beta
\end{equation*}
as soon as 
\begin{equation*}
C \geq \frac{1}{\sqrt{d}}  
\left(\sigma^{k+1} \sqrt{2\log\frac{\tau}{\alpha}}
 + \zeta^{k+1}_{j_{0}} \sqrt{-2\log \beta}\right).
\end{equation*}

\paragraph{Procedure P2}Let $C$ be a real such that $\nabla^{k+1}p_{j_{0}} \leq -C$, and let $\zeta^{k+1}_{j_0} = \sqrt{ A^{k T}_{j_0}\Gamma A^{k+1}_{j_0}}$. We have:
\begin{eqnarray*}
 \P\left(  \widehat{\cS}_{\alpha}^{k} \geq
  0 \vert D=d\right) & = &
\P\left( \min_{0\leq j \leq \widehat{\tau}-1}
 \left\{\sqrt{d} \nabla^{k+1}{f}_{j} - \nu_{\widehat{u}^{k+1}_{\alpha}} \sqrt{ A^{k+1 T}_{j}\widehat{\Gamma} A^{k+1}_{j}}\right\} \geq
  0\right) \\
 &\leq &
\P\left( \sqrt{d} (\nabla^{k+1}f_{j_{0}} -\nabla^{k+1}p_{j_{0}})\geq
  \nu_{\widehat{u}^{k+1}_{\alpha}} \sqrt{ A^{k+1 T}_{j_0}\widehat{\Gamma} A^{k+1}_{j_0}}+C\sqrt{d}\right) \\
  &\leq &
\P\left( \sqrt{d} A^{k+1 T}_{j_{0}} \Z \geq \nu_{u^{k+1}_{\alpha}} \zeta^{k+1}_{j_0}+C
  \sqrt{d} \right) + o(1)
\end{eqnarray*}
Applying the classical Tchebychev inequality, we get
\begin{equation}
\label{aux15juin3}
 \P\left( \sqrt{d} A^{k+1 T}_{j_{0}} \Z \geq \nu_{u^{k+1}_{\alpha}} \zeta^{k+1}_{j_0}+C
  \sqrt{d} \right)  \leq 
\exp\left(-\frac{(\nu_{u^{k+1}_{\alpha}} \zeta^{k+1}_{j_0}+C
  \sqrt{d})^{2}}{2 (\zeta^{k+1}_{j_{0}})^{2}}\right).
\end{equation}
Let us remark that applying Formula (\ref{aux15juin4}) to the case where  $C=0$,  we get
\begin{equation*}
\alpha\leq  \tau \max_j\;u_{\alpha}^{k}
\end{equation*}
then $u_{\alpha}^{k+1}\geq \alpha/\tau$ and
\begin{equation}
\label{maj.q.eq2}
\nu_{u_{\alpha}^{k+1}} \geq \nu_{\alpha/\tau} \geq -
\sqrt{2\log\frac{\tau}{\alpha}}.
\end{equation}
Considering inequalites given at Equations \eref{aux15juin3} and \eref{maj.q.eq2}, we get that
\begin{equation*}
 \P_{\bar{H}^{k+1}(C)}\left( \sqrt{d} A^{k+1 T}_{j_{0}} \Z \geq
  \nu_{u^{k+1}_{\alpha}} \zeta^{k+1}_{j_0}+C
  \sqrt{d} \right) \leq \beta
\end{equation*}
as soon as 
\begin{equation*}
C \geq \frac{1}{\sqrt{d}}  
\left(\sqrt{2\log\frac{\tau}{\alpha}} 
 +  \sqrt{-2\log \beta}\right)\zeta^{k+1}_{j_0}.
\end{equation*}

\subsection{Proof of Theorem~\ref{kChap.th}}

If $k=1$,
\begin{eqnarray*}
\P\left( \widehat{k}_{\alpha} =0 \right) &=&
\P\left( H^{1}\mbox{  is rejected} \right) \leq \alpha + o(1)
\end{eqnarray*}
Let us now consider the case where $k\geq 2$
\begin{eqnarray*}
\P\left( \widehat{k}_{\alpha} \leq k-1 \right) &=&
\P\left( \exists \ell, 1\leq\ell\leq k-1, \forall m\leq\ell,  H^m \mbox{ is
    not rejected and } H^{\ell+1} \mbox{ is rejected} \right) \\
& \leq & \P\left( \exists \ell, 1\leq\ell \leq k-1, H^{\ell+1} \mbox{ is rejected}\right) \\
& \leq & \P\left( H^{k}\mbox{  is rejected} \right) +
\mathbb{I}_{\left(k \geq 3\right)} \sum_{\ell=1}^{k-2}\P\left( H^{\ell+1} \mbox{ is rejected} \right)
\end{eqnarray*}
Thanks to Theorem~\ref{TestkLevel.th}, taking $\beta =
1/(k-2)\sqrt{d}$ if $k\geq 3$, we
get the first part of the Theorem.

For the second part of the theorem
\begin{eqnarray*}
\P\left( \widehat{k}_{\alpha} \geq k+1 \right) &=& \sum_{\ell=k+1}^{k_{\max}} 
\P\left( \widehat{k}_{\alpha} = \ell \right) \\
&=& \sum_{\ell=k+1}^{k_{\max}} 
\P\left( \forall m, 1 \leq m \leq \ell, H^{m}\mbox{ is not rejected and } H^{\ell+1} \mbox{ is rejected} \right) \\
& \leq & \sum_{\ell=k+1}^{k_{\max}} 
\P\left( \forall m, k+1 \leq m \leq \ell, H^{m}\mbox{ is not rejected } \right)\\
& \leq & (k_{\max}-k-1)
\P\left(  H^{k+1}\mbox{ is not rejected } \right).
\end{eqnarray*}

\subsection{\label{biasVar.st}Bias and variance}

\begin{eqnarray*}
\widehat{N}^{k}  & = &  D  - \sum_{h=1}^{k} (-1)^{h} C_{k}^{h}S_{h} \\
& = & D +S_{0} -\nabla^{k} S_{0} \\
\E\left(\widehat{N}^{k} \right)& = &  N - N \nabla^{k} p_{0}
\end{eqnarray*}

Let $\alpha_{k,h} = 1-(-1)^{h} C_{k}^{h}$
\begin{eqnarray*}
\V\left( \widehat{N}^{k}/\sqrt{N}\right) &= &
\sum_{h=1}^{k} \alpha_{k,h}^{2}  p_{h}(1-p_{h}) 
- \sum_{h_{1}\neq h_{2}}\alpha_{k,h_{1}}\alpha_{k,h_{2}}
p_{h_{1}} p_{h_{2}}
-2\sum_{h=1}^{k} \alpha_{k,h}
p_{h}p_{\geq k+1}
+p_{\geq k+1}(1-p_{\geq k+1}) \\ &=&
\sum_{h=1}^{k} \alpha_{k,h}^{2}p_{h} - \sum_{h=1}^{k}
\alpha_{k,h}^{2}p_{h}^{2} - \left(\sum_{h=1}^{k}
  \alpha_{k,h}p_{h}\right)^{2}
+ \sum_{h=1}^{k} \alpha_{k,h}^{2}p_{h}^{2} \\ 
&& -2\sum_{h=1}^{k} \alpha_{k,h}p_{h}p_{\geq k+1} +p_{\geq k+1} -
p_{\geq k+1}^{2} \\ &=&
\sum_{h=1}^{k} \alpha_{k,h}^{2}p_{h}  - \left(\sum_{h=1}^{k}
  \alpha_{k,h}p_{h} +p_{\geq k+1}\right)^{2}
  +p_{\geq k+1} \\ &=&
\sum_{h=1}^{k} \left(1-(-1)^{h} C_{k}^{h}\right)^{2}p_{h}  - \left(\sum_{h=1}^{k}
\left(  1-(-1)^{h} C_{k}^{h}\right)p_{h} +p_{\geq k+1}\right)^{2}
  +p_{\geq k+1}\\ &=&
\sum_{h=1}^{k} \left(1-2(-1)^{h} C_{k}^{h}+\left(C_{k}^{h}\right)^{2}\right)p_{h}  - \left(\sum_{h=1}^{k}p_{h}+p_{\geq k+1}
-\sum_{h=1}^{k}(-1)^{h} C_{k}^{h}p_{h} \right)^{2}
  +p_{\geq k+1}\\ &=&
1-p_{0} -2\sum_{h=1}^{k} (-1)^{h} C_{k}^{h}p_{h} +
\sum_{h=1}^{k}\left(C_{k}^{h}\right)^{2}p_{h} -
\left(1-p_{0} -\sum_{h=1}^{k} (-1)^{h} C_{k}^{h}p_{h}\right)^{2} \\
&=&
p_{0}(1-p_{0})  -2\sum_{h=1}^{k} (-1)^{h} C_{k}^{h}p_{h} +
\sum_{h=1}^{k}\left(C_{k}^{h}\right)^{2}p_{h}
+2(1-p_{0})\sum_{h=1}^{k} (-1)^{h} C_{k}^{h}p_{h}
\\ && -\left(\sum_{h=1}^{k} (-1)^{h} C_{k}^{h}p_{h}\right)^{2}\\ &=&
p_{0}(1-p_{0})  -2p_{0}\sum_{h=1}^{k} (-1)^{h} C_{k}^{h}p_{h} +
\sum_{h=1}^{k}\left(C_{k}^{h}\right)^{2}p_{h}
-\left(\sum_{h=1}^{k} (-1)^{h} C_{k}^{h}p_{h}\right)^{2}
\\ &=&
p_{0}-\left(\sum_{h=1}^{k} (-1)^{h} C_{k}^{h}p_{h} +
  p_{0}\right)^{2} +
\sum_{h=1}^{k}\left(C_{k}^{h}\right)^{2}p_{h}\\ &=&
p_{0}+
\sum_{h=1}^{k}\left(C_{k}^{h}\right)^{2}p_{h} - \left(\nabla^{k}p_{0}\right)^{2}.
\end{eqnarray*}

\section*{References}

\bibliographystyle{abbrvnat}
\bibliography{biblioARTICLE}

\end{document}